\documentclass[11pt]{article}
 \usepackage[greek,english]{babel}
 \usepackage[iso-8859-7]{inputenc}
 \usepackage{epsfig}
\begin{document}

\begin{center} 
{\bf \Large  On the dynamics of Extrasolar Planetary Systems under dissipation. Migration of planets}
\end{center}
\begin{center}
{John D. Hadjidemetriou and George Voyatzis\\ Department of Physics, Aristotle University
of Thessaloniki, 54124, Greece.\\ e-mail: hadjidem@auth.gr, voyatzis@auth.gr}
\end{center}

{\bf Abstract:} {\small We study the dynamics of planetary systems with two planets moving in the same plane, when frictional forces act on the two planets, in addition to the gravitational forces. The model of the general three-body problem is used. Different laws of friction are considered. The topology of the phase space is essential in understanding the evolution of the system. The topology is determined by the families of stable and unstable periodic orbits, both symmetric and non symmetric. It is along the stable families, or close to them, that the planets migrate when dissipative forces act. At the critical points where the stability along the family changes, there is a bifurcation of a new family of stable periodic orbits and the migration process changes route and follows the new stable family up to large eccentricities or to a chaotic region. We consider both resonant and non resonant planetary systems. The 2/1, 3/1 and 3/2 resonances are studied. The migration to larger or smaller eccentricities depends on the particular law of friction. Also, in some cases the semimajor axes increase and in other cases they are stabilized. For particular laws of friction and for special values of the parameters of the frictional forces, it is possible to have partially stationary solutions, where the eccentricities and the semimajor axes are fixed.} 

\section{Introduction}
Since the last decade of the 20th century it is known that there are 340 observed extrasolar planetary systems with more than 400 planets and some of them have two or more planets (Schneider, October 2009).
In many cases of multiplanetary systems, two planets are in mean motion resonance, (e.g.  HD 82943, GJ 876, HD128311, 55CnC$_{b,c}$). Some of these systems have large eccentricities and are evidently stable, since they exist in nature. Thus the role of the resonances in the evolution of a planetary system should play an important role either for its long term stability or its migration to the present position. 

There are different approaches to the study of the long term dynamical stability of a resonant planetary system  and on the mechanisms that stabilize the system or generate chaotic motion and instability (e.g. Beaug\'e et al, 2003; Ferraz-Mello et al, 2006;  Marzari et al, 2006; Callegari et al, 2006; Baluev, 2008; Gozdziewski et al, 2008; Beaug\'e et al, 2008; Schward et al, 2009).  In these papers different methods have been applied, as the averaging method, direct numerical integrations of orbits, or various numerical methods which provide indicators for the exponential growth of nearby orbits. In this way the regions where stable motion exists have been detected, in the orbital elements space. Another method to study the dynamics of a planetary system is to find the basic families of periodic orbits and their stability. It is known that the periodic orbits play a dominant role in understanding the dynamics of a system, because they determine critically the structure of the phase space. In this way, we can detect the regions where stable librations could exist (Hadjidemetriou, 2006; Voyatzis, 2008). These are the regions close to the stable periodic orbits, where a real planetary system could be trapped. The periodic orbits with nonzero eccentricities of the planets, either stable or unstable, are associated with mean motion resonances between the planets. Also, work on these lines are by Hadjidemetriou (2002), Psychoyos and Hadjidemetriou (2005), Voyatzis and Hadjidemetriou (2005, 2006). A review on these issues is given in Hadjidemetriou (2008). 

It is believed that the existence of a planetary system in a region of long term stability is the result of a dynamical process that forces the planets to migrate after their formation (see e.g. Tsiganis et al, 2005; Morbidelli et al, 2007).  Friction, due to the interaction between the newly formed planets and the protoplanetary nebula, before this latter is dissolved, is the main source of non conservative forces that generate a {\it migration} of the planets. Also tidal friction due to the body deformation also produces non conservative forces that affects both the orbit and the spin evolution (Ferraz-Mello 2008). Non conservative forces may explain the large eccentricities that are observed in some planetary systems or the very small planetary distances from the sun. 

The introduction of non conservative forces destroy the symplectic structure of a purely gravitational system and the phase space volume is not conserved. Consequently the overall dynamics and the topology of the phase space change critically. A particular review and relative literature can be found in Contopoulos (2002).   

Numerical simulations using hydrodynamical models (e.g. Nelson and Papaloizou, 2003a,b) or dissipative restricted three body models (Beaug\'e and Ferraz-Mello, 1993; Gomes, 1995; Haghighipour, 1999)  showed that planetary evolution leads to a trapping of planets in a mean motion resonance. Such evolution follows  paths of resonant stable configurations (Lee and Peal 2002; Papaloizou 2003; Lee, 2004; Sandor and Kley, 2006). Such paths are proved to be characteristic curves of resonant corotations (Ferraz-Mello et al, 2003; Beaug\'e et al 2006; Zhou et al, 2008), which actually coinside with families of periodic orbits in a rotating frame (Voyatzis and Hadjidemetriou 2005, 2006). The present work aims to study the close association of a possible evolution of a planetary system under dissipative forces with the families of periodic orbits of the conservative model. As we shall see in the following, the families of periodic orbits play a crucial role on the evolution of the system, in phase space, under any dissipation law and the knowledge of the families of periodic orbits is necessary in understanding the migration process.      

We consider, additionally to the gravitational forces, the action of the Stokes non conservative force 
\begin{equation}
\vec{R}=-10^{-\nu}(\vec{v}-\vec{v_c}),\;\;\nu>0
\label{diss1}
\end{equation}
where the velocity $\vec{v_c}$ is the Keplerian circular velocity at the distance $r$ from the Sun, given by
$$
\vec v_c=\sqrt{\frac {Gm_0}{r}}\:\vec{e_\theta}. 
$$
The vector $\vec e_\theta$ is the unit vector normal to the radius vector $\vec r$ and $m_0$ is the mass of the sun. The Stokes dissipation law (\ref{diss1}) implies that the nebula that generates the dissipation rotates differentially with a Keplerian circular velocity $\vec v_c$ at any distance $r$. We also consider a variation of the Stokes force given by 
\begin{equation}
\vec{R}=-10^{-\nu}(\vec{v}-\vec{v_c}/r^{1/2}),\;\; \nu>0.
\label{diss2}
\end{equation}
This law differs from the law given in Equation (\ref{diss1}), in that the circular velocity is multiplied by $1/r^{1/2}$. This means that the  velocity of rotation of the protoplanetary nebula varies with the distance from the sun. It is larger than $\vec v_c$ close to the sun and goes slowly to zero as $r\rightarrow \infty$. We remark that $\vec{R}$ is applied to both planets and not only to the outer planet as e.g. in the studies of Lee and Peal (2002), Beaug\'e et al (2006). 

The model we use is the {\it general three body problem}, for planar motion, with the sun, $S$, and the two planets, $P_1$ (inner) and $P_2$ (outer), as the three bodies of mass $m_0$, $m_1$ and $m_2$, respectively, normalized such that $m_0+m_1+m_2=1$ . We assume that the center of mass of the whole system is at rest with respect to an {\it inertial} frame $XOY$. We have four degrees of freedom, for planar motion, with generalized variables $X_1,\:Y_1,\:X_2,\:Y_2$. We define next, a {\it rotating} frame of reference $xOy$, whose $x$-axis is the line $S-P_1$, with origin at the center of mass of these two bodies, where $S$ is the Sun and $P_1$ the inner planet. This is a non uniformly rotating frame, where the planet $P_1$ moves  on the $x$-axis and the planet $P_2$ moves in the $xy$ plane. As generalized coordinates in this rotating frame we can select $x_1,\:x_2,\:y_2$ and $\theta$, where $\theta$ is the angle between the $Ox$ axis and a fixed direction in the inertial frame. It turns out that the angle $\theta$ is ignorable, so we have three degrees of freedom and the study of the system is restricted in the rotating frame $xOy$, with variables $x_1,\:x_2,\:y_2$ (Hadjidemetriou, 1975). 

The evolution of the planetary system can be studied by computing the Poincar\'e map on a surface of section. By this method we reduce the dimensions of the phase space, without losing the generality of the problem. In the present study, we consider the surface of section
\[y_2=0 \:\:(\dot y_2>0),\]
in the rotating frame $xOy$. 

It can be shown (see e.g. Voyatzis and Hadjidemetriou 2005, 2006) that families of symmetric or asymmetric periodic orbits exist in the {\it rotating} frame $xOy$. In particular, in a symmetric periodic orbit the planet $P_2$ intersects perpendicularly the $x$-axis while the planet $P_1$ is temporarily at rest on the $x$-axis. Consequently, the non zero initial conditions of a symmetric periodic orbit are $x_{10}$, $x_{20}$, $\dot y_{20}$, and a family of symmetric periodic orbits is represented by a smooth curve in the space $\Pi_3=\{x_{10},x_{20},\dot y_{20}\}$ of initial conditions. An asymmetric periodic orbit is represented by a smooth curve in a five dimensional space of initial conditions e.g. $\Pi_5=\{x_{10}, x_{20},\dot{x}_{10},\dot x_{20}, \dot y_{20}\}$. In the following we shall present a family by its projection on the coordinate plane $x_{10}x_{20}$ or in the space $e_1e_2$ of the osculating eccentricities at $t=0$. When we plot symmetric periodic orbits in the space $e_1e_2$ we use the convention $e_i>0$ or $e_i<0$ when the planet is at aphelion or at perihelion, respectively.

In the conservative system the stable fixed points of the Poincar\'e map correspond to periodic orbits in the rotating frame. If in the equations of motion we consider dissipative forces, in addition to the gravitational forces, the energy integral does not exist. The stable fixed points are now stable limit cycles, to which the system is attracted in some cases. Chaotic attractors may also exist and correspond to a set of points (as $t\rightarrow \infty$), which are irregularly distributed in the section.

\section{The evolution of a non resonant orbit under dissipation}
In this section we study the evolution of a planetary system which starts from a non-resonant orbit and evolves under the influence of the dissipation laws (\ref{diss1}) or (\ref{diss2}).  

\begin{figure}
\begin{center}
\includegraphics[width=6cm,height=6cm]{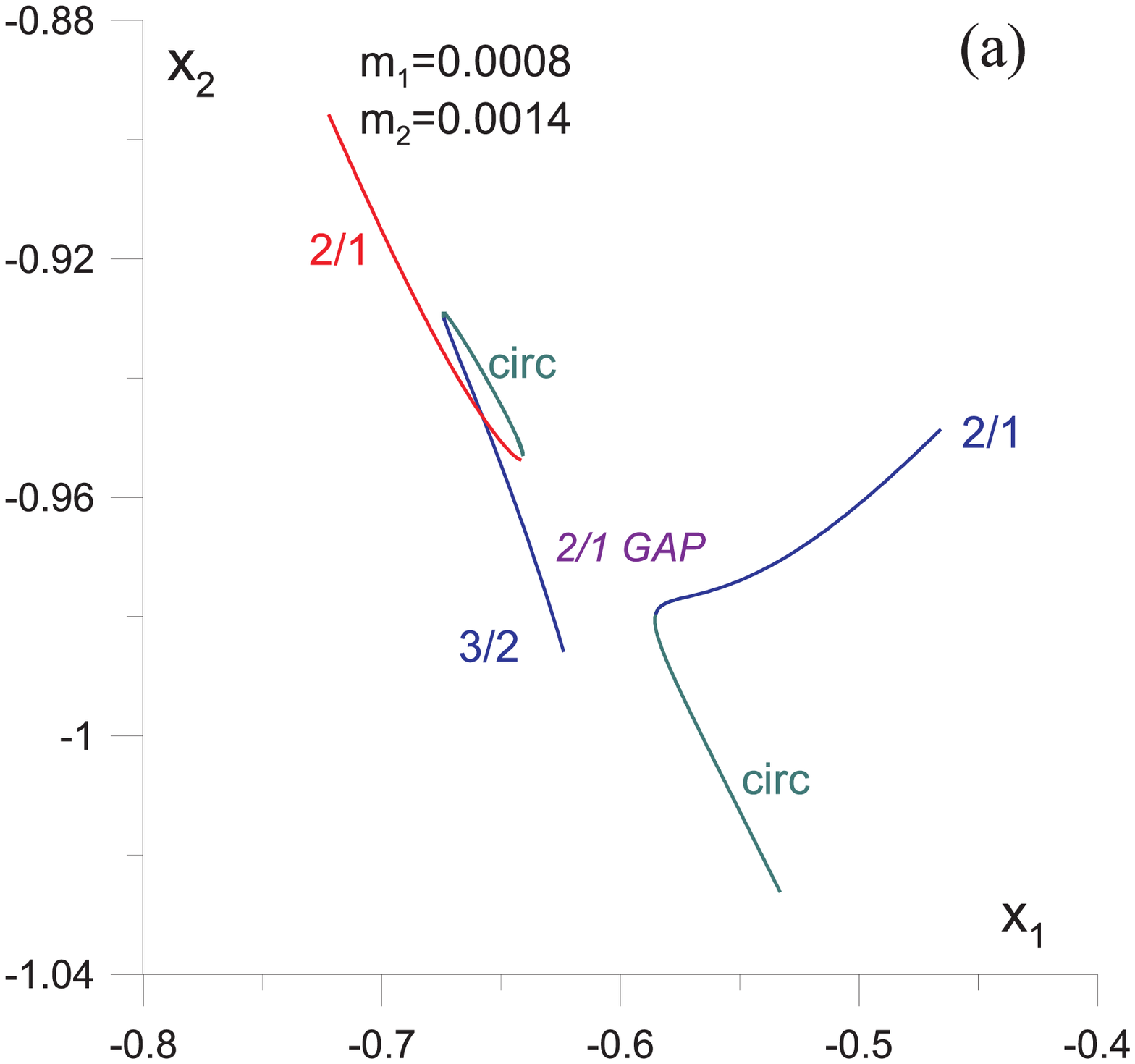}\hspace{1cm}
\includegraphics[width=6cm,height=6cm]{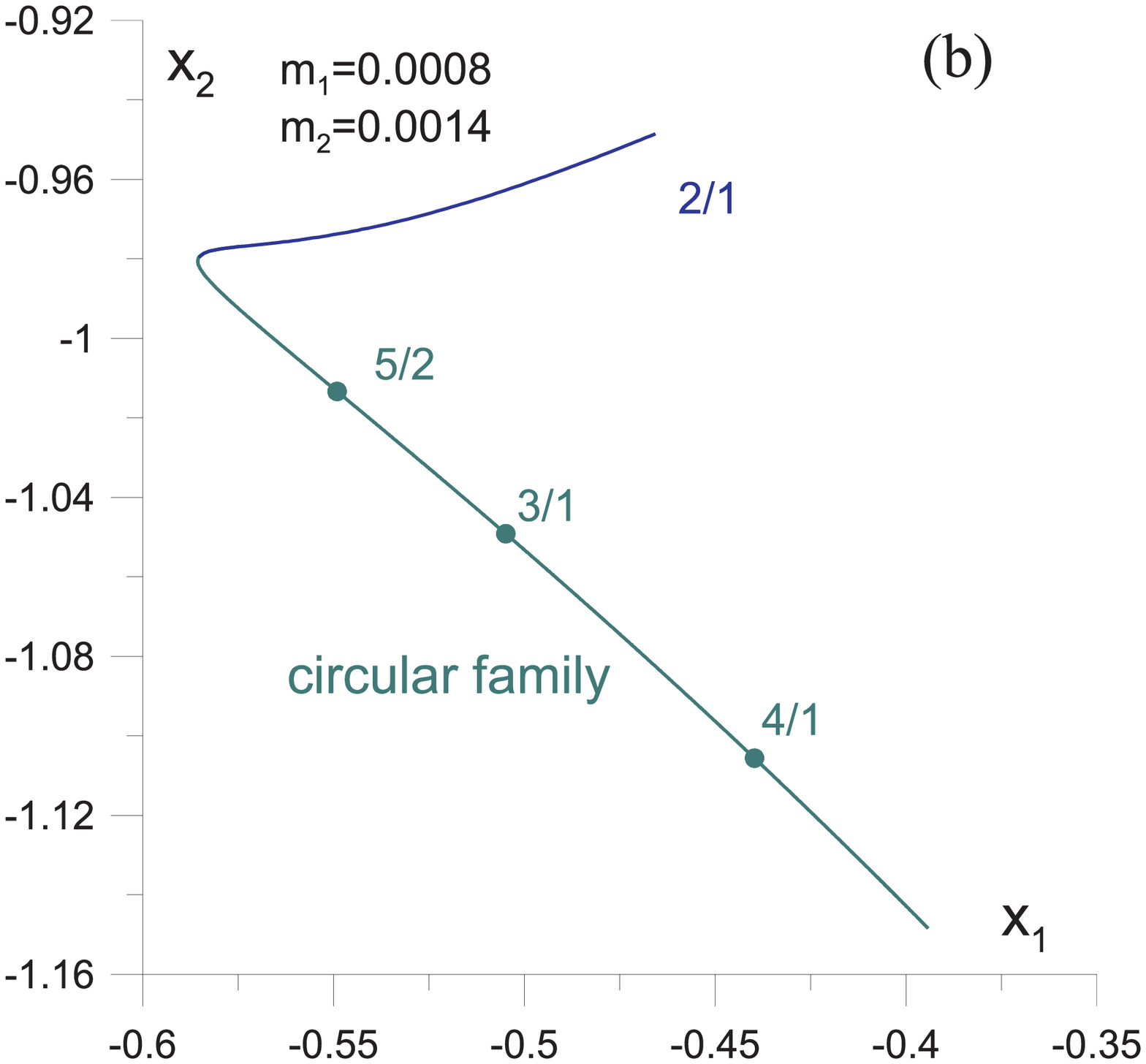}
\end{center}
\caption{(a) The family of circular (non resonant) periodic orbits and the families of 2/1 and 3/2 resonant periodic orbits close to the 2/1 gap. (b) The non resonant circular family and the continuation to the 2/1 resonant family after the 2/1 gap. The circular orbits at the 4/1, 3/1 and 5/2 resonances are indicated.}
\label{families}
\end{figure}

In Fig. \ref{families}a we present the family of circular orbits and the families of 2/1 and 3/2 resonant periodic orbits, in their projection on the $x_{10}x_{20}$ plane. The family of circular orbits brakes at the 2/1 resonance, where a gap appears and we have a bifurcation of two families of 2/1 resonant elliptic periodic orbits. The same phenomenon is repeated on the circular family at the 3/2 resonance, where a 3/2 resonant family of elliptic periodic orbits bifurcates (Hadjidemetriou, 2006). In Fig. \ref{families}b we present a detail of Fig. \ref{families}a, where we show the family of non resonant circular orbits and the bifurcation of one branch of the 2/1 resonant family. On the circular family the 4/1, 3/1 and 5/2 resonant circular orbits are indicated. These orbits are for the masses $m_0=0.9978$, $m_1=0.0008$ and $m_2=0.0014$. The family of circular orbits in Fig. \ref{families}b is stable except close to the 3/1 resonance. The 2/1 resonant family in this figure is stable, for $m_1<m_2$, but a small unstable region appears when $m_1>m_2$. These will be discussed in the following. 

\subsection{The dissipation law $\vec{R}=-10^{-\nu}(\vec{v}-\vec{v_c})$}

We start our study by considering a non resonant planetary system with orbital eccentricities $e_1=e_{10}$, $e_2=e_{20}$ and a fixed value of the ratio of the planetary frequencies $n_1/n_2$. The initial values for the phases $M$ (mean motion) and $\omega$ (longitude of pericenter) for both planets are taken equal to $0^\circ$ or $180^0$, but it was verified that the evolution is similar for any other initial values. 

\subsubsection{Case I: Starting from $n_1/n_2\approx 2.8$}

\begin{figure}
\begin{center}
\includegraphics[width=6cm,height=6cm]{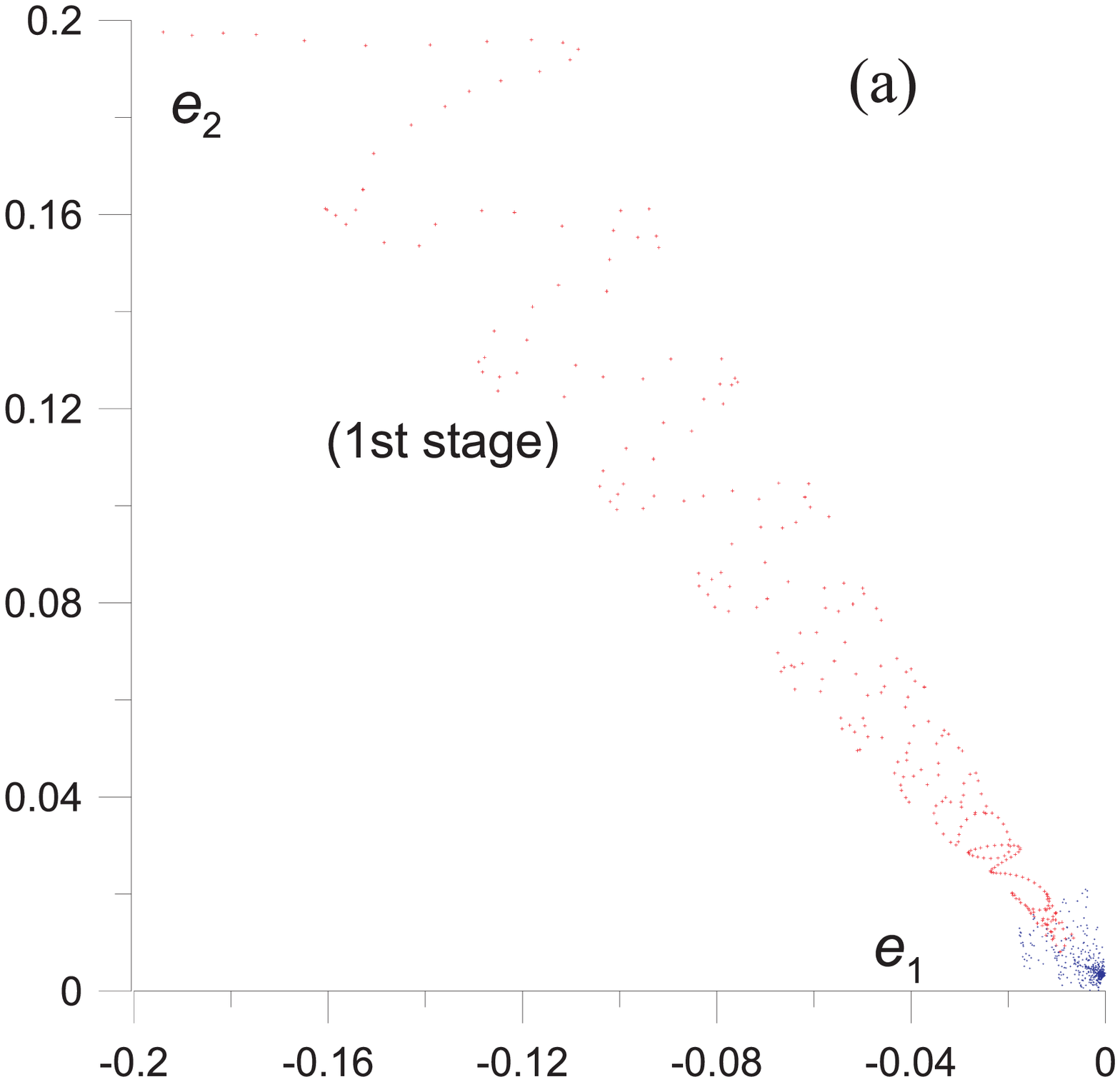}\hspace{1cm}
\includegraphics[width=6cm,height=6cm]{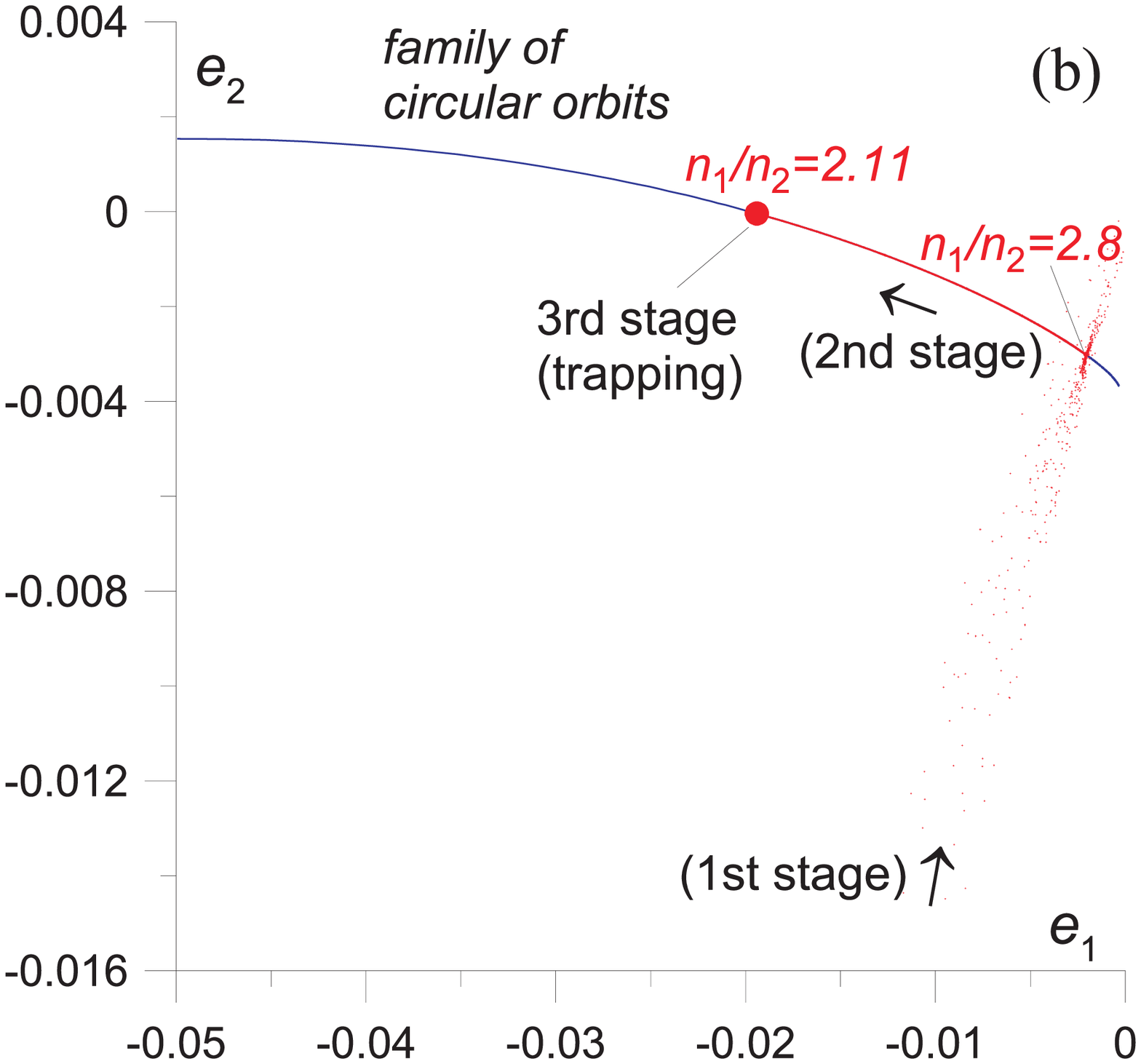}
\end{center}
\caption{The evolution of the system (in the eccentricity space) under the dissipation law (\ref{diss1}), starting from $n_1/n_2=2.8$ and $e_1=0.20,\:e_2=0.20$. (a) The system is attracted to an orbit on the circular family. (b) After it reaches the circular family, the system moves smoothly on the circular family until it reaches an orbit close to the 2/1 resonance and is trapped there.}
\label{e12-28}
\end{figure}

\begin{figure}
\begin{center}
\includegraphics[width=8cm]{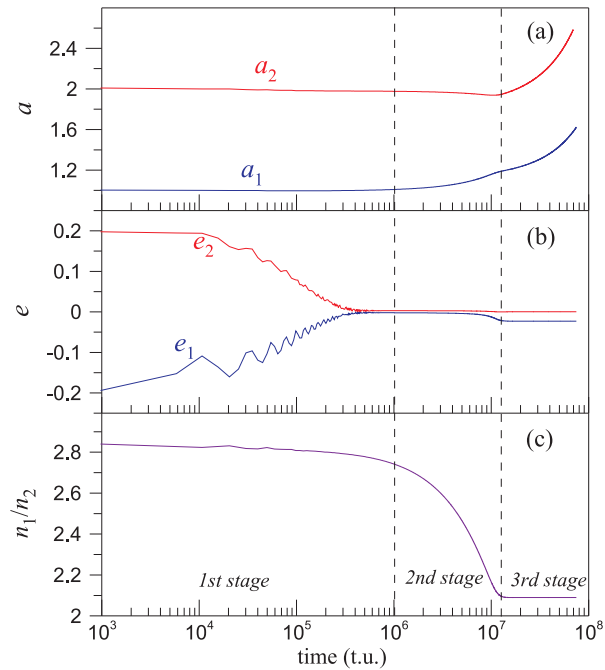}
\end{center}
\caption{The evolution of the planetary orbital elements associated to the system evolution presented in Fig. \ref{e12-28}. The distinction between the three stages of Fig. \ref{e12-28}b is clearly seen. (a) Evolution of the semimajor axes. (b) Evolution of the eccentricities. (c) Evolution of the ratio $n_1/n_2$.}
\label{t28}
\end{figure}

We consider a planetary system with initial eccentricities $e_{10}=0.20$, $e_{20}=0.20$ and $a_{10}=1.0$, $a_{20}=2.0$, corresponding to $n_1/n_2\approx 2.8$, and for the angles we have taken the initial values $M_1=0$, $M_2=0$, $\omega_1=0$, $\omega_2=180^0$. This is a non resonant orbit, whose position could be represented in the Fig. \ref{families}b by a point far from the circular family. The evolution of this system in the eccentricity space, under the dissipation law (\ref{diss1}) with $\nu=5$, is presented in Fig. \ref{e12-28}. In panel (a) we present the first stage of the evolution, where the system is attracted to the family of circular orbits, keeping the ratio $n_1/n_2$ close to its initial value ($n_1/n_2\approx 2.8$). This motion is quite irregular. In panel (b) we present the second stage of the evolution, namely, after the trapping of the system in a periodic orbit of the circular family ($e_i\approx 0$). The system follows a route {\it along} the circular family, which is also presented, and is attracted to an orbit close to the 2/1 resonance (third stage), just before the 2/1 gap (see Fig. \ref{families}b). 

In Fig. \ref{t28} we present the evolution of the system in time. We use use a logarithmic scale for the time to show better the three stages of evolution.  During the first stage the semimajor axes, and, subsequently, the ratio $n_1/n_2$ are almost constant. Up to this point, there is an agreement with the results of a first order analysis given by Beaug\'e et al (2006). During the second stage the variation of semimajor axes causes a decrease of the ratio $n_1/n_2$ up to the resonance 2/1. From that point on the ratio $n_1/n_2$ remains almost constant and the semimajor axes increase (third stage).  The eccentricities (panel (b)) decrease at first, while the system moves towards the circular family . The eccentricities are finally trapped close to zero (second and third stage of evolution). 

\subsubsection{Case II: $n_1/n_2\approx 3.5$}
\begin{figure}
\begin{center}
\includegraphics[width=4.5cm]{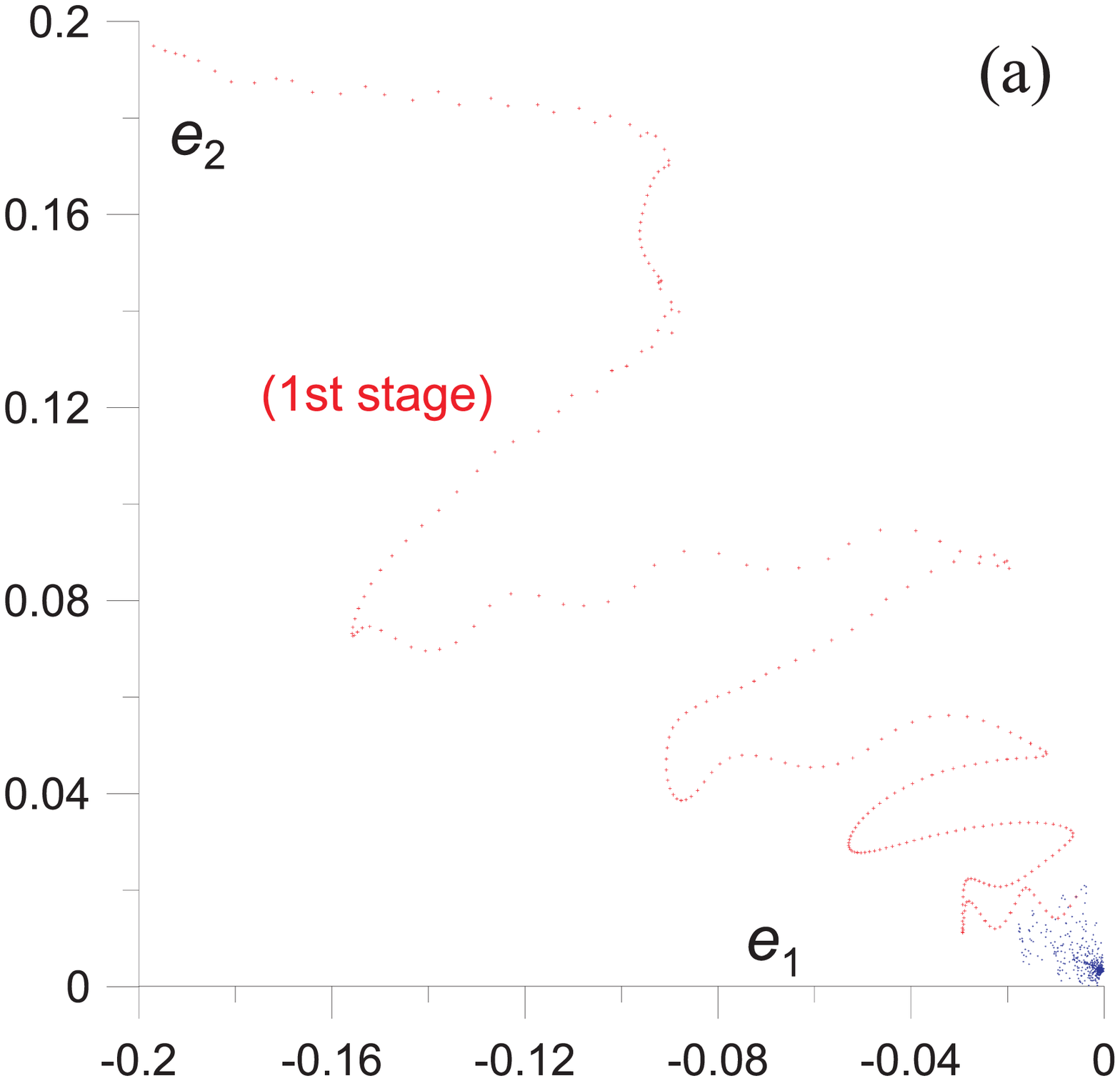}\hspace{1cm}
\includegraphics[width=4.5cm]{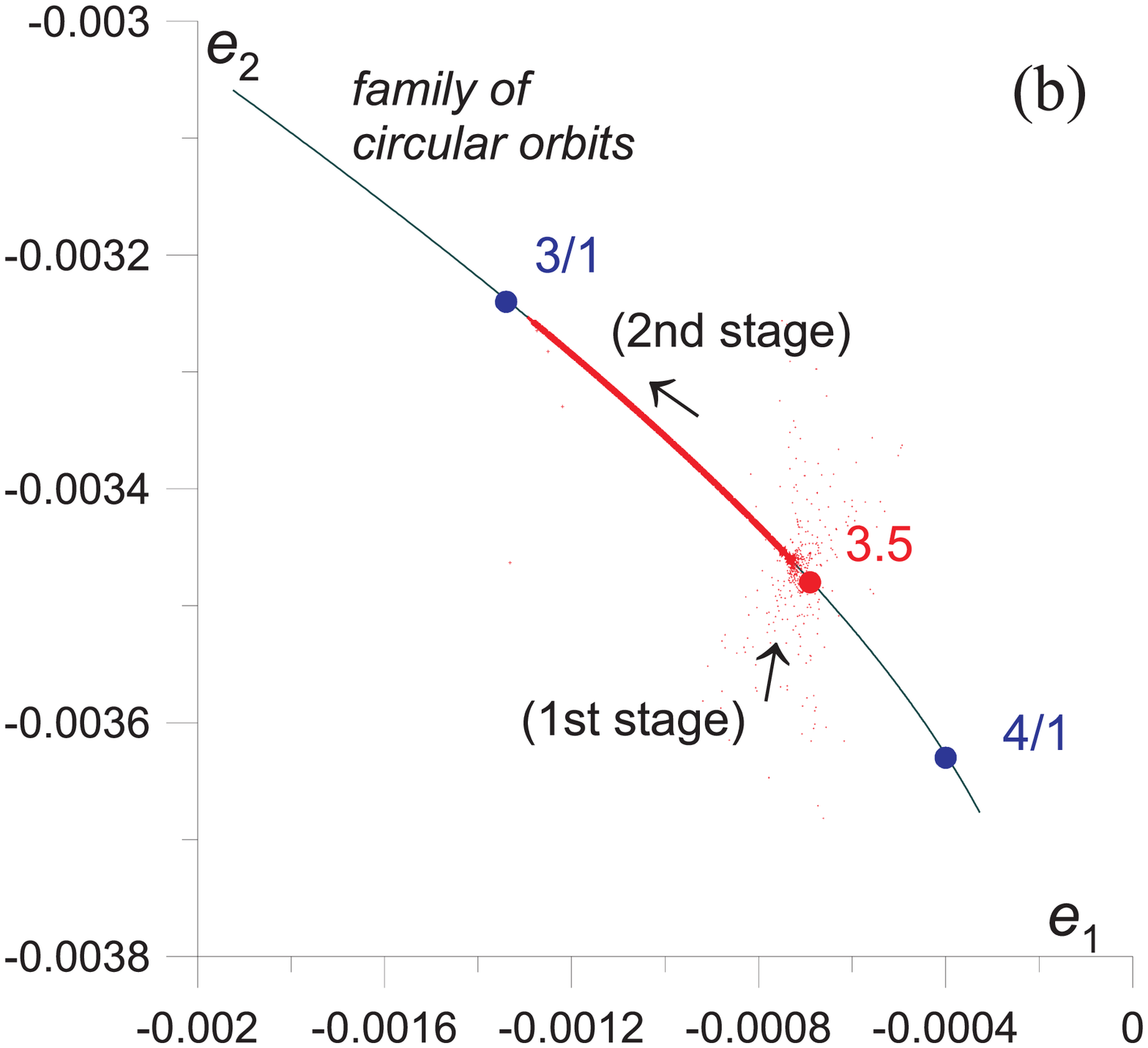}\hspace{1cm}
\includegraphics[width=4.5cm]{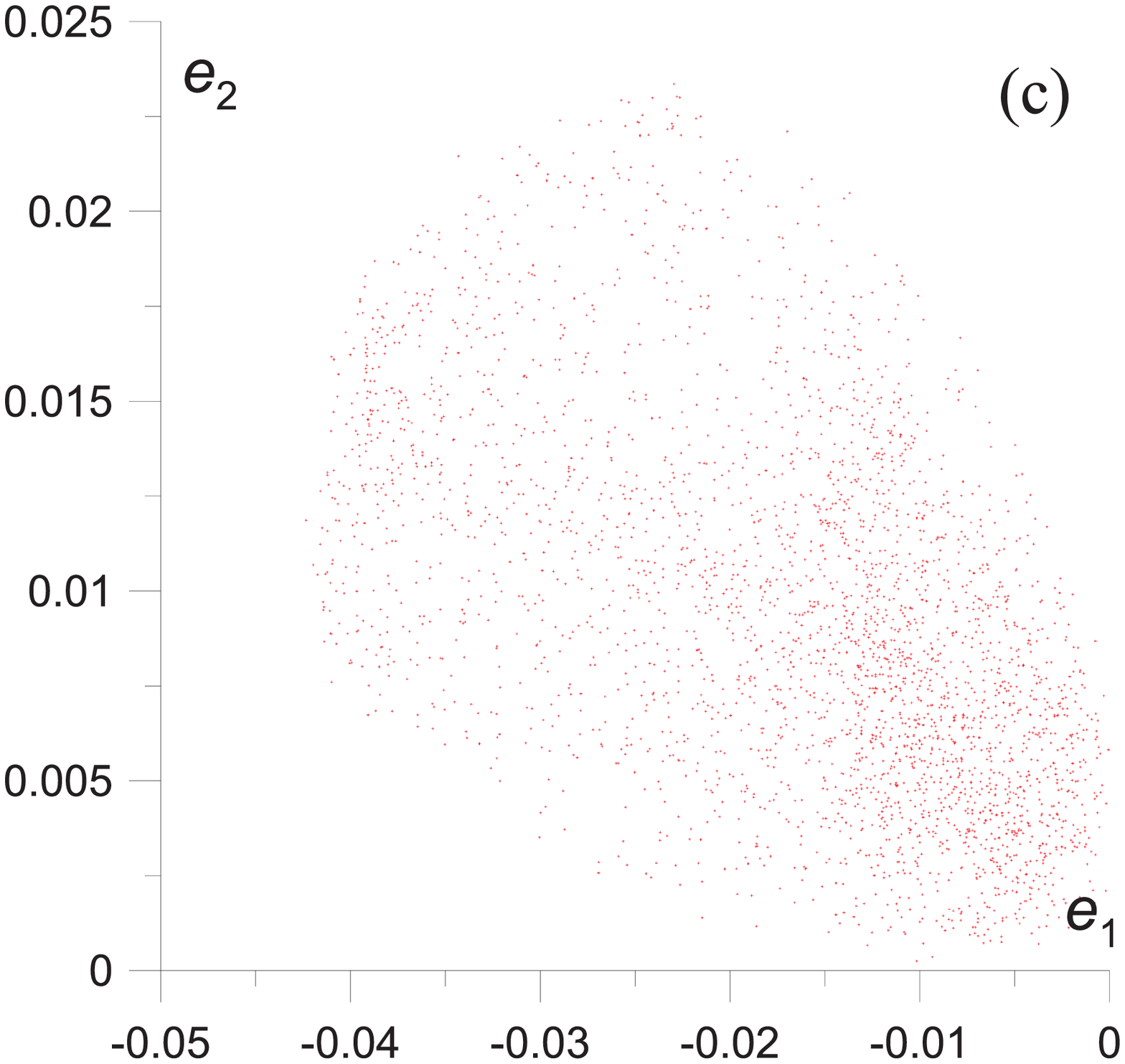}
\end{center}
\caption{(a) The first stage of the evolution of the system (in the eccentricity space) under the dissipation law $\vec{R}=-10^{-5}(\vec{v}-\vec{v_c})$, 
starting from $n_1/n_2\approx 3.5$ and $e_1=0.20,\:e_2=0.20$. The system is attracted to an orbit on the circular family. (b) The second stage of the evolution : the system moves smoothly on the circular family. (c) The system is trapped in a chaotic attractor close to the circular orbit at the 3/1 resonance.}
\label{e12-35}
\end{figure}
\begin{figure}
\begin{center}
\includegraphics[width=8cm]{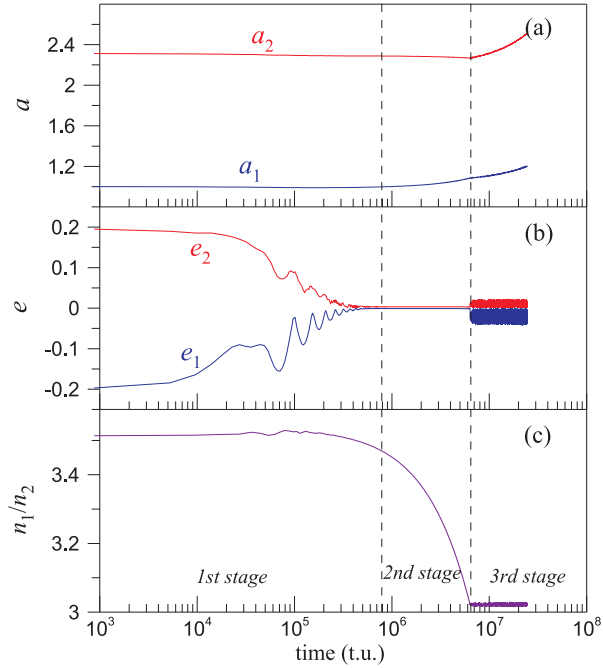} 
\end{center}
\caption{The evolution of the elements of the planetary orbits presented in figure \ref{e12-35}. The distinction between the three stages of Fig. \ref{e12-35}b is clearly seen. (a) Evolution of the semimajor axes. The semimajor axes increase steadily after the system is trapped close to the 3/1 resonance. (b) The evolution of the eccentricities. They remain close to zero even after the trapping to the chaotic attractor. c) The evolution of the ratio $n_1/n_2$.}
\label{t35}
\end{figure}

We consider a planetary system with initial conditions $e_{10}=0.20$, $e_{20}=0.20$, $a_{10}=1.0$, $a_{20}=2.3$, corresponding to $n_1/n_2\approx 3.5$, and angles $M_1=0$, $M_2=0$, $\omega_1=0$, $\omega_2=180^0$.  We study the evolution of this system in the same way as in the previous section, under the Stoke's dissipation $\vec{R}=-10^{-5}(\vec{v}-\vec{v_c})$. The evolution of this system in the eccentricity space is presented in Fig. \ref{e12-35}. In panel (a) we present the first stage and in panel (b) the second stage of the evolution, similarly to the previous case. But now, the system is finally attracted to a chaotic attractor close to the 3/1 resonance, possibly associated with the unstable region at the 3/1 resonance (panel (c)). The evolution of the semimajor axes, the eccentricities and the mean motion ratio is presented in Fig. \ref{t35} and shows similar behaviour as in the previous case. We remark the irregular oscillations of the eccentricities near to zero after the system is attracted in the chaotic attractor.   

\begin{figure}
\begin{center}
\includegraphics[width=16cm]{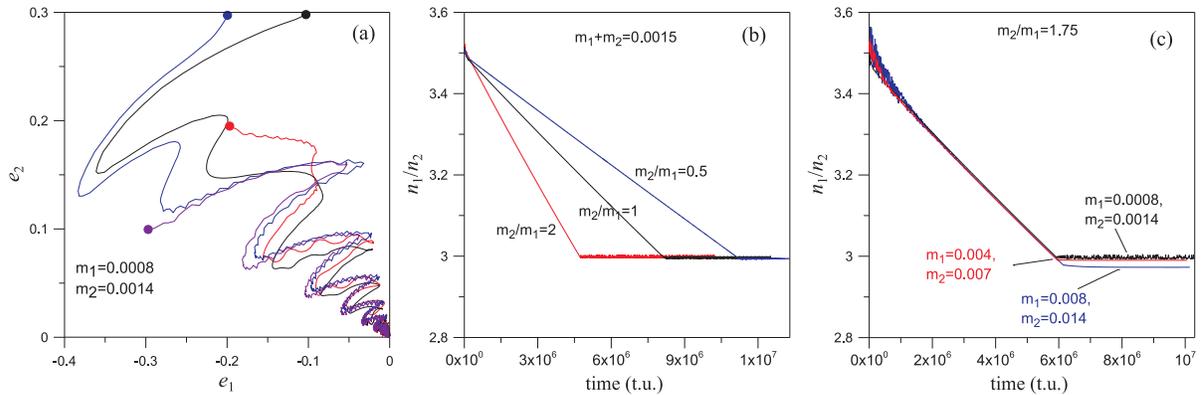}
\end{center}
\caption{(a) The evolution of the system, as in Fig. \ref{e12-35}, for different initial eccentricities. (b) The evolution of $n_1/n_2$ for different planetary mass ratio values and $m_1+m_2=0.0015$. (c) The evolution of $n_1/n_2$ for different mass values and $m_2/m_1=1.75$.}
\label{e12-35-all}
\end{figure}

In order to check the universality of the previous results, we repeated the study presented in Fig. \ref{e12-35}, for different initial conditions in the eccentricities, keeping the initial ratio $n_1/n_2\approx 3.5$. The results are presented in Fig. \ref{e12-35-all}a: In the whole region in the eccentricity space $0<e_i<0.30$, the system is attracted first to the same periodic orbit on the circular family (at $n_1/n_2\approx 3.5$), and after this the evolution is the same as that described in Fig. \ref{e12-35}. We see that this region in the eccentricity space is the {\it basin of attraction} towards the chaotic attractor at the 3/1 resonance of Fig. \ref{e12-35}c. 

We also repeated the study for different values of the planetary masses and verified that the evolution is qualitatively the same, as that described above. In figure \ref{e12-35-all}b we present the evolution of the mean motion ratio for different mass ratios (but for the same total planetary mass). Indeed we obtain the same feature of the decreasing $n_1/n_2$ but the rate of decreasing, and, subsequently, the moment of the final capture in the resonance, depends on the mass ratio. Instead, if we change both planetary masses, preserving the mass ratio, the decreasing rate of $n_1/n_2$ is not affected (see Fig. \ref{e12-35-all}c).

\subsection{The dissipation law $\vec{R}=-10^{-\nu}(\vec{v}-\vec{v_c}/r^{1/2})$}
\begin{figure}[t]
\begin{center}
\includegraphics[width=6.5cm]{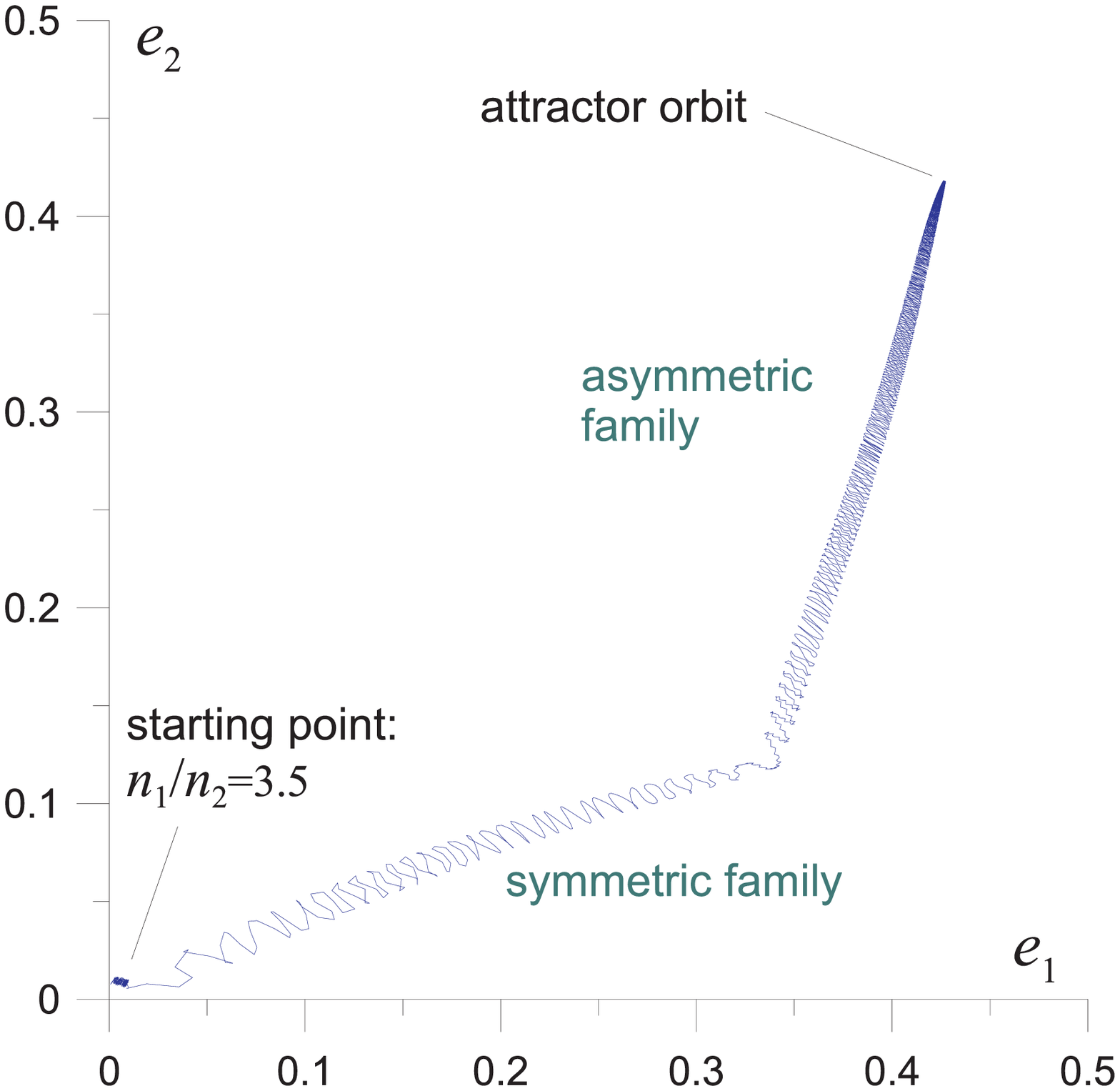}\hspace{1cm}
\includegraphics[width=6.5cm]{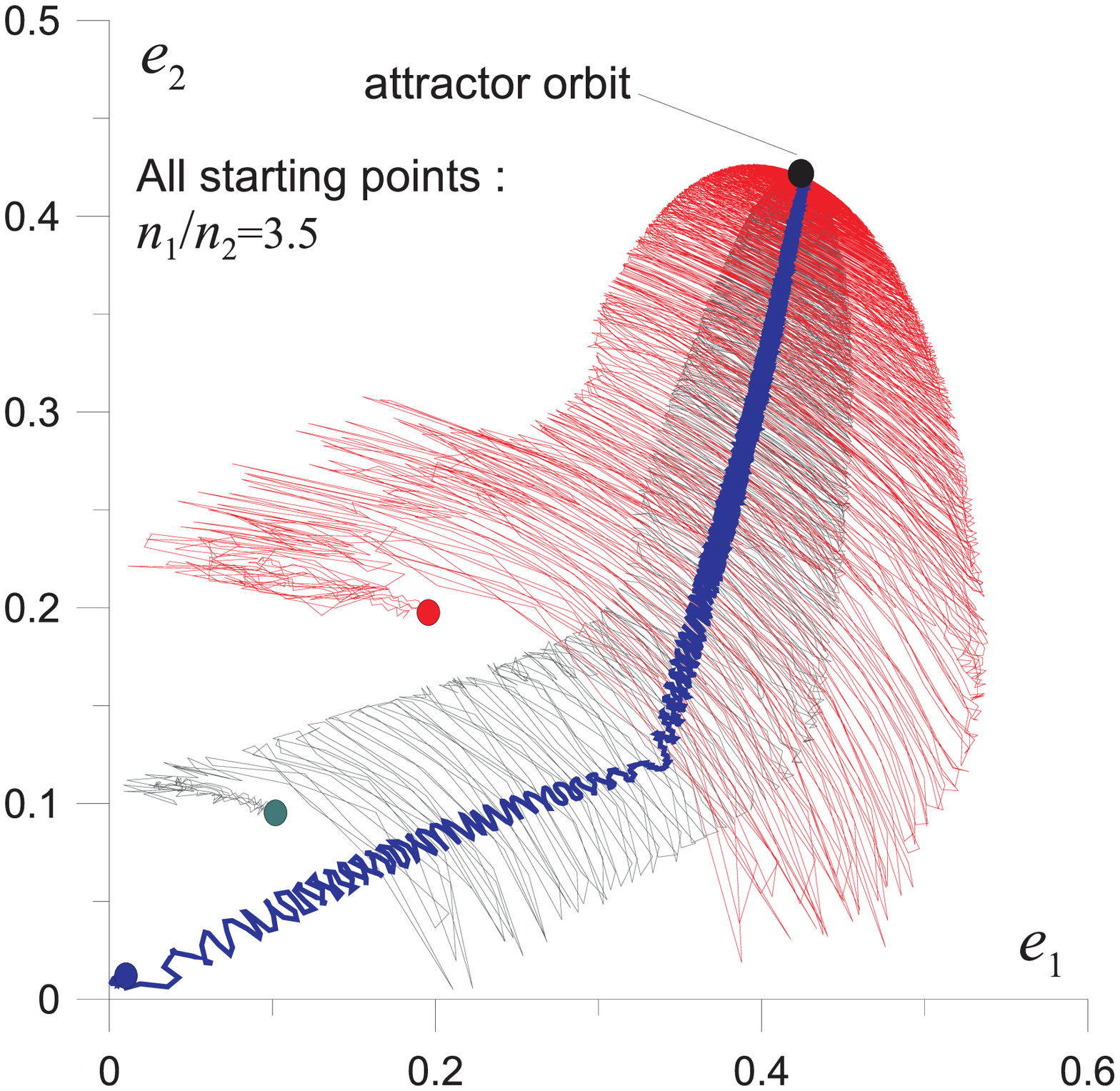}
\end{center}
\caption{The evolution of the system (in the eccentricity space) under the dissipation law $\vec{R}=-10^{-6}(\vec{v}-\vec{v_c}/r^{1/2})$, 
starting from $n_1/n_2=3.5$. (a) Starting eccentricities: $e_{10}=0.01$ $e_{20}=0.01$. (b) The same evolution as in panel (a) and, additionally, for $e_{10}=0.10$ $e_{20}=0.10$ and $e_{10}=0.20$, $e_{20}=0.20$ . In all cases the system is attracted to the same asymmetric periodic orbit.}
\label{e12-4}
\end{figure}
\begin{figure}
\begin{center}
\includegraphics[width=16cm]{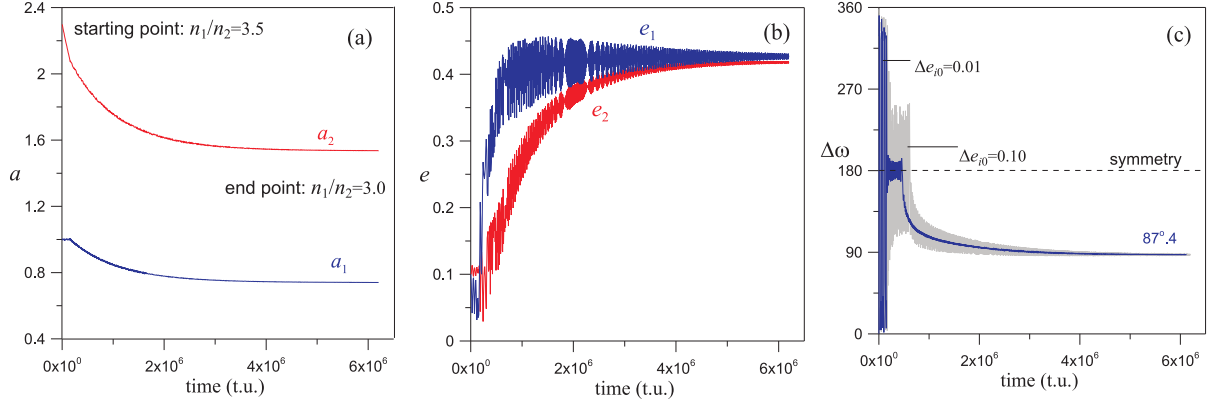}
\end{center}
\caption{The evolution of Fig. \ref{e12-4}a, in the elements of the orbit. (a) The evolution of the semimajor axes. (b) The evolution of the eccentricities. (c) The evolution of the planetary apsidal difference $\Delta\omega$ starting close ($\Delta e_{i0}=0.01$) and far ($\Delta e_{i0}=0.1$) from the periodic orbit; the transition from the symmetric to the asymmetric part of the 3/1 family is clearly seen.}
\label{t35-4}
\end{figure}

In this paragraph we study the evolution of a system under the dissipation law (\ref{diss2}).
We work as in the previous section, starting from a planetary system with elements $0.01\leq e_{i0} \leq 0.2$, $i=1,2$, and $a_{10}=1.0$, $a_{20}=2.3$, corresponding to $n_1/n_2\approx 3.5$. For the angles we have taken the initial values $M_1=0$, $M_2=0$, $\omega_1=0$, $\omega_2=180^0$, but, as it has been mentioned, the evolution is similar for any other values of the angles. A presentation of the evolution is given in figures \ref{e12-4} and \ref{t35-4}. 

In panel (a) of Fig. \ref{e12-4} we present the evolution for $e_{10}=0.01$ and $e_{20}=0.01$. This means that the system is a non resonant system with $n_1/n_2\approx 3.5$, with very small eccentricities, almost on the family of circular orbits mentioned before and is close to the 3/1 resonant orbit on this family. We note that a small unstable region appears on the circular family, at the 3/1 resonance, and from the critical points at the ends of this unstable region there bifurcate four symmetric families of 3/1 resonant elliptic periodic orbits, one stable and the rest unstable (Voyatzis and Hadjidemetriou 2006; Voyatzis, 2008). The ratio $n_1/n_2$ decreases rapidly up to the 3/1 resonance and from that point on the system follows the path along the {\it stable} symmetric family of 3/1 resonant periodic orbits. However, this family becomes unstable at a certain point, and another 3/1 resonant elliptic family of asymmetric periodic orbits bifurcates from this critical point. The system changes route and follows the stable asymmetric family, until it is trapped at a periodic orbit on this family, with relatively large eccentricities. In panel (b) of Fig. \ref{e12-4} we present the same evolution starting from different initial eccentricities. All orbits are attracted to the same periodic orbit as above. In Fig. \ref{t35-4} we present the evolution of the system in the elements of the orbit. In panel (a) we show the evolution of the semimajor axes. They decrease at first, and then they remain constant. In panel (b) it is shown that the eccentricities increase at first and then they are stabilized to the constant values $e_1=0.419$, $e_2=0.427$. Consequently, the evolution is attracted to a particular periodic orbit which is asymmetric as it is clearly indicated by the evolution of the angle $\Delta \omega$ of the line of apsides of the two planetary orbits. Initially the angle $\Delta \omega$ oscillates around the value $180^\circ$ that corresponds to the motion on the symmetric family and finally tends to the value $\Delta \omega=87^\circ.4$, which indicates clearly the asymmetry of the orbit.

\begin{figure}
\begin{center}
\includegraphics[width=12cm]{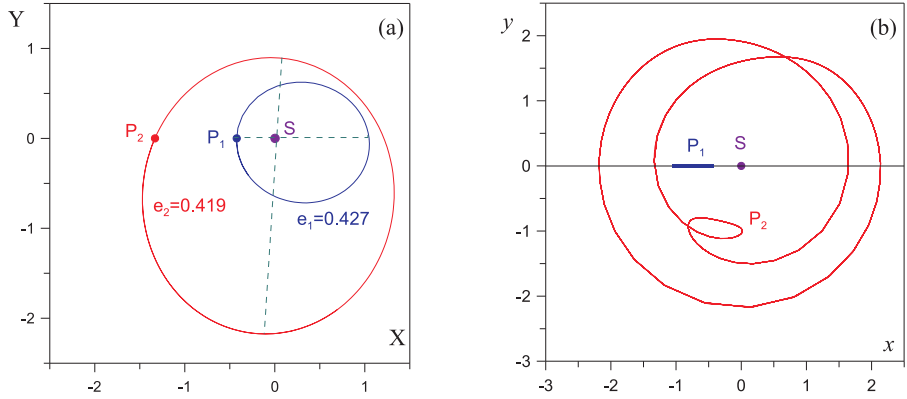}
\end{center}
\caption{The asymmetric periodic orbit to which the system is trapped in the evolution of Fig. \ref{e12-4}, in (a) the inertial and (b) the rotating frame.}
\label{orb}
\end{figure}

This asymmetric orbit, which is the attractor of the evolution, is shown in Fig. \ref{orb}. In order to stop the planetary migration process and obtain an attractor orbit, Lee and Peale (2002) artificially introduce to the system a sufficient amount of eccentricity damping. In our case the migration stops naturally if the dissipation law (\ref{diss2}) is taken into account. 

\begin{figure}
\begin{center}
\includegraphics[width=6.5cm]{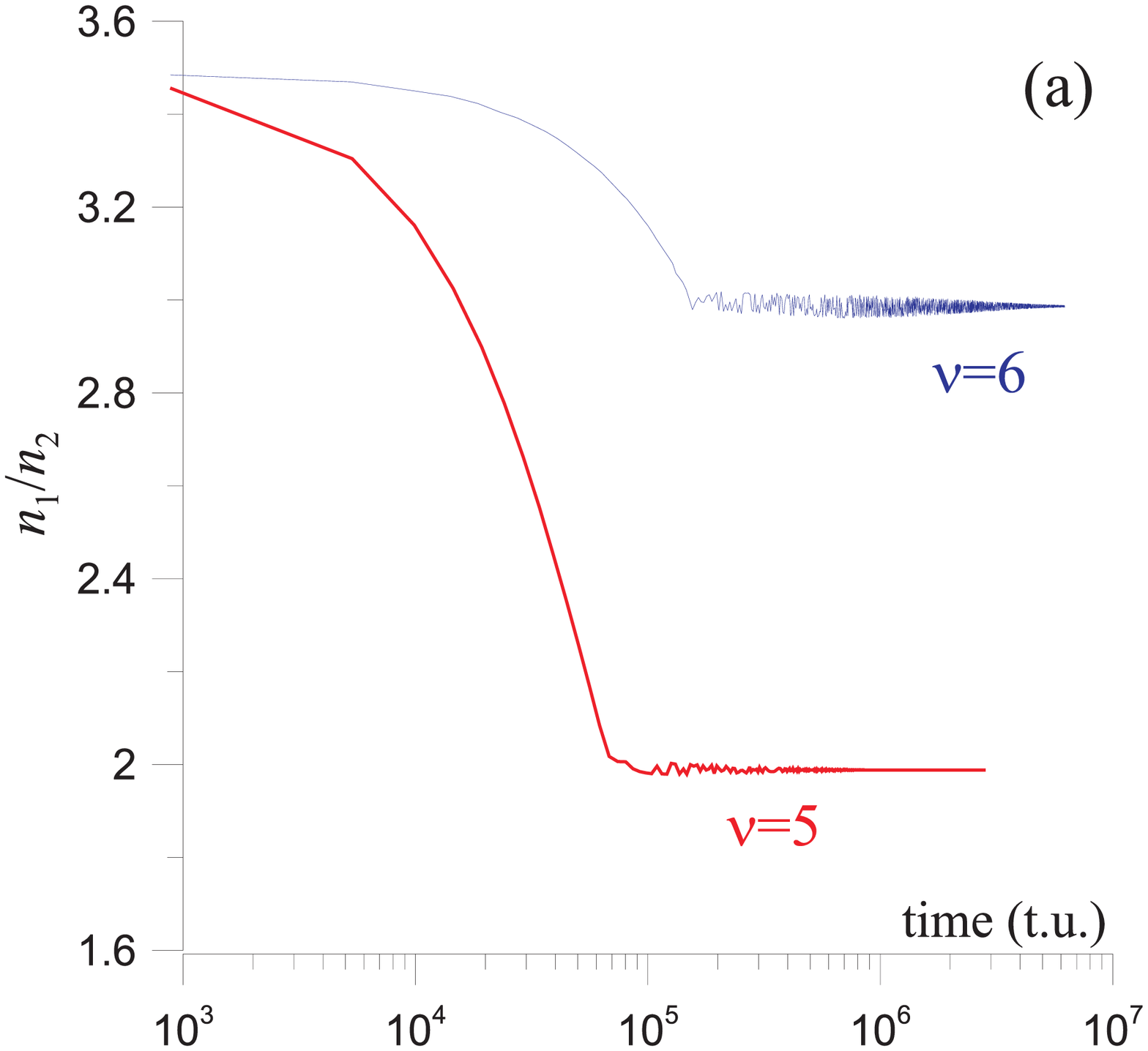}\hspace{1.0cm}
\includegraphics[width=6cm]{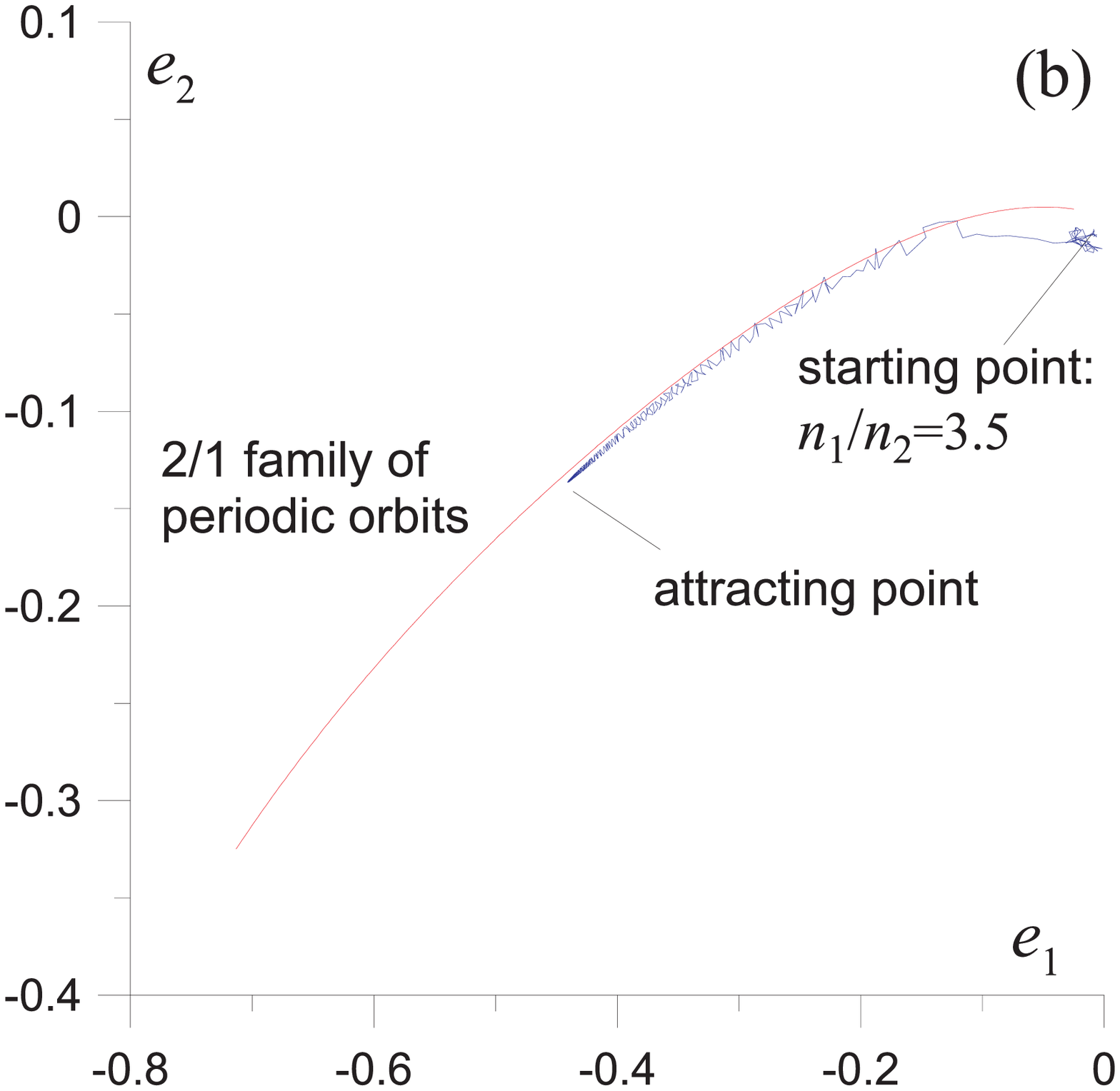}
\end{center}
\caption{(a) The evolution of Fig. \ref{e12-4}, in the ratio $n_1/n_2$ starting from $n_1/n_2=3.5$. For $\nu=6$ the system is trapped in the 3/1 resonance but for $\nu=5$ the system is trapped in the 2/1 resonance. (b) The trapping in the 2/1 resonance of panel (a) in the eccentricity plane.}
\label{tres-35}
\end{figure}

Finally, we present in Fig. \ref{tres-35}a the evolution of the ratio $n_1/n_2$, under the dissipation law $\vec{R}=-10^{-\nu}(\vec{v}-\vec{v_c}/r^{1/2})$, for $\nu=5$ and $\nu=6$. We found that in almost all cases the system is trapped at the 3/1 resonance, as shown in Fig. \ref{e12-4}. We found only one special case, corresponding to $\nu=5$ and $e_{10}\leq 0.01$, $e_{20}\leq 0.01$, where the system is not trapped to the 3/1 resonance, as the ratio $n_1/n_2$ decreases, but goes up to the 2/1 gap. From that point on, it is trapped and moves {\it on} the stable family of 2/1 resonant elliptic periodic orbits, until finally it is attracted to a 2/1 resonant periodic orbit on the 2/1 family, with large eccentricities. This is shown in Fig. \ref{tres-35}b. 

\section{The evolution close to resonant orbits under dissipation:\\ 2/1 resonance}

\subsection{The dissipation law $\vec{R}=-10^{-\nu}(\vec{v}-\vec{v_c})$} \label{Sec21Res}
In this section we study the evolution of a planetary system under the dissipation law (\ref{diss1}). We consider orbits which start in a region close to the families of 2/1 resonant symmetric periodic orbits. Since the structure of the families of periodic orbits is essentially different between the cases $m_1<m_2$ and $m_1>m_2$, we study both cases separately. A detailed study of the families of periodic orbits at the 2/1 resonance is given in Voyatzis and Hadjidemetriou (2005), Beaug\'e et al (2006), Voyatzis et al (2009). We mention that similar dynamics with that presented below is obtained for the 3/1 resonance too. 

\subsubsection{The inner planet has a smaller mass than the outer planet.} 

\begin{figure}
\begin{center}
\includegraphics[width=6.5cm]{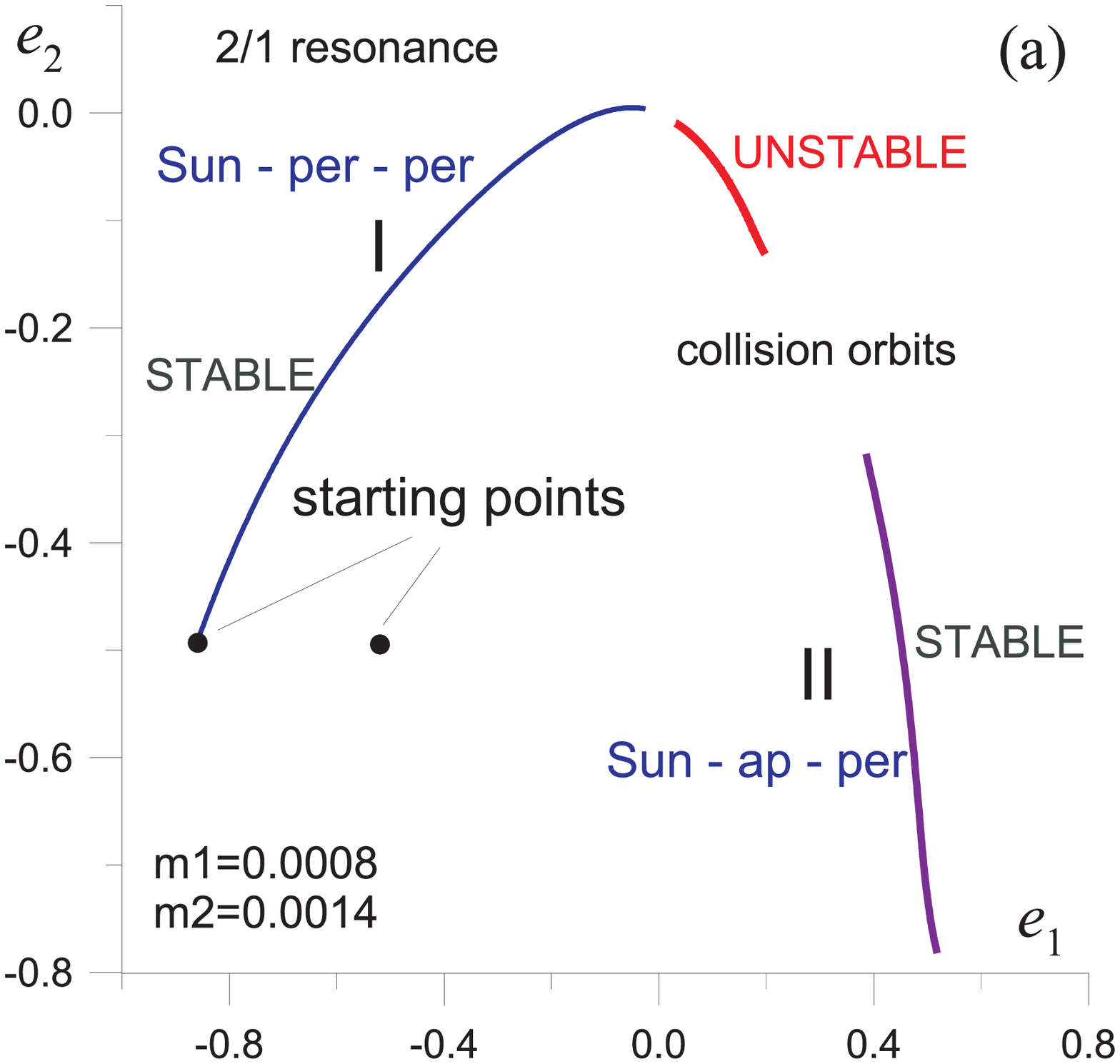}\hspace{0.5cm}
\includegraphics[width=6.5cm]{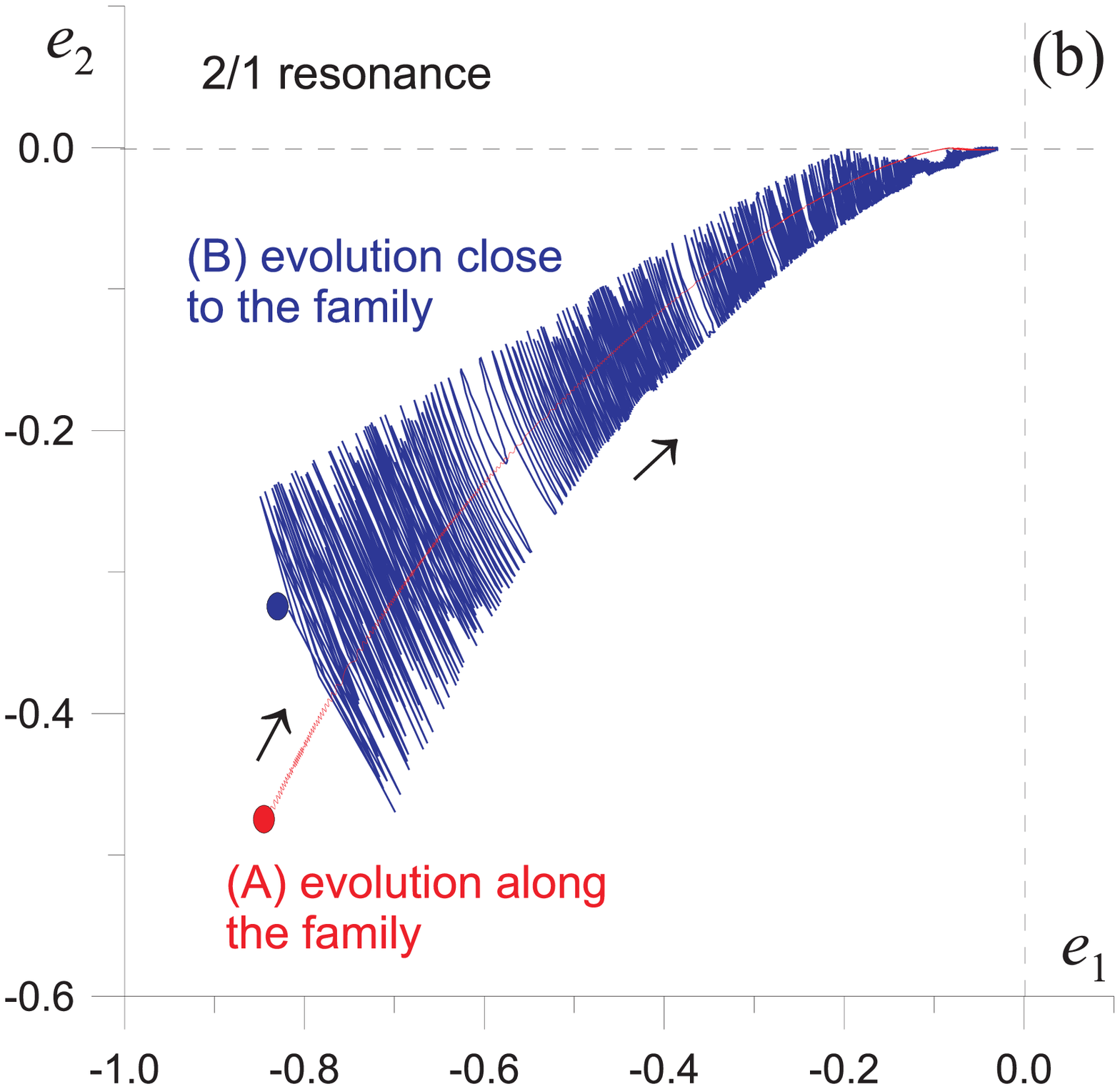}
\end{center}
\caption{(a) The two families of 2/1 resonant periodic orbits and the two starting points for the evolution of the system under dissipation. (b) The evolution of the elements of the planetary orbits under the dissipation law $\vec{R}=-10^{-7}(\vec{v}-\vec{v_c})$, starting  on the family I of the 2/1 resonant periodic orbits (case A) and far from it (case B).}
\label{fam21-1}
\end{figure}

We used the mass values $m_1=0.0008$ and $m_2=0.0014$. We start in a region close to the 2/1 resonant family presented in Fig. \ref{families}a. There are two resonant symmetric families, separated by the 2/1 gap. In Fig. \ref{fam21-1}a we present these two families in the eccentricity space $e_1e_2$, denoted by I and II.  In family I both planets are at perihelion ($\Delta\omega=0$) and in family II the inner planet is at perihelion and the outer at aphelion ($\Delta\omega=180^0$). The 2/1 gap of Fig. \ref{families}a appears in this diagram as a small gap at $e_1\approx 0$, $e_2\approx 0$. The family I is all stable. The family II starts as unstable, for small eccentricities, until a gap appears, corresponding to close collision orbits between the planets. After this collision area, the family II is stable up to large eccentricities. 

We study the evolution of the system  by starting from the two points indicated in Fig. \ref{fam21-1}a. The first point is {\it on} the family, with high eccentricities ($e_1=0.80$, $e_2=0.40$). The second point is obtained by shifting the value $e_1$ by 0.14 and keeping the value of $e_2$ the same.  We indicate the two evolution cases as $A$ and $B$. The evolution, for $\nu=7$ is shown in Fig. \ref{fam21-1}b.  In the case A, the system moves along this family, with {\it decreasing} eccentricities (in absolute values), until it is trapped to a periodic orbit with almost zero eccentricities, close to the 2/1 resonance. In the evolution of orbit B, the system also follows a path along the family, which acts as a guiding line, with oscillations around the family that decrease in amplitude and tends to the same zero eccentricity orbit as in the case A.

\begin{figure}
\begin{center}
\includegraphics[width=16cm]{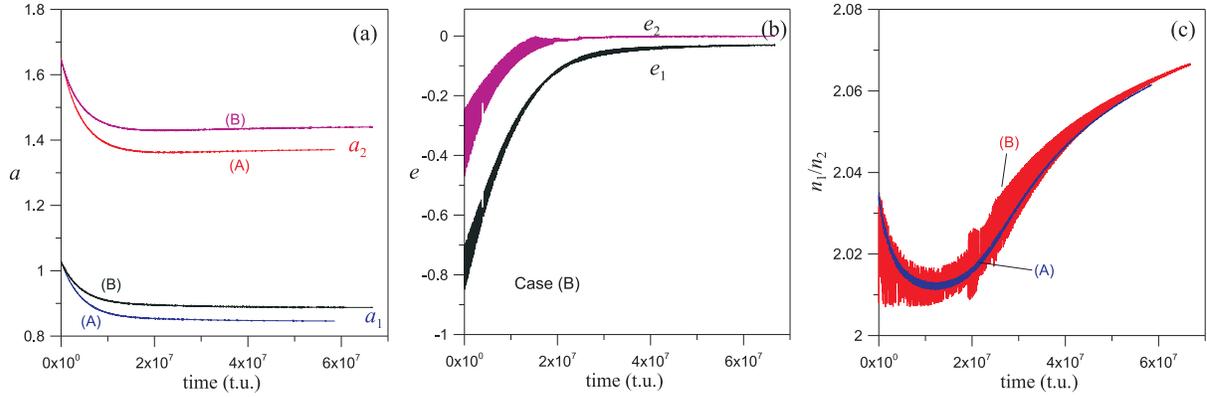}
\end{center}
\caption{The evolution of Fig. \ref{fam21-1}b, in the elements of the orbit. Both the evolution {\it along} the family (case A) and {\it close} to the family (case B) are shown. (a) The evolution of the semimajor axes. (b) The evolution of the eccentricities (only case B is shown). (c) The evolution of $n_1/n_2$.}
\label{t-21-1}
\end{figure}

In Fig. \ref{t-21-1} we present the evolution of Fig. \ref{fam21-1}b, in the elements of the orbit. In panel (a) we show the evolution of the semimajor axes of the two planets. They decrease at first and then are stabilized, as the system is trapped at the 2/1 resonant periodic orbit with zero eccentricities. In panel (b) we show the evolution of the eccentricities. They decrease (in absolute values) at first and stay at small values after the trapping to the zero eccentricity orbit mentioned above. The ratio $n_1/n_2$ stays close to the 2/1 value (panel (c)).

\begin{figure}
\begin{center}
\includegraphics[width=10cm]{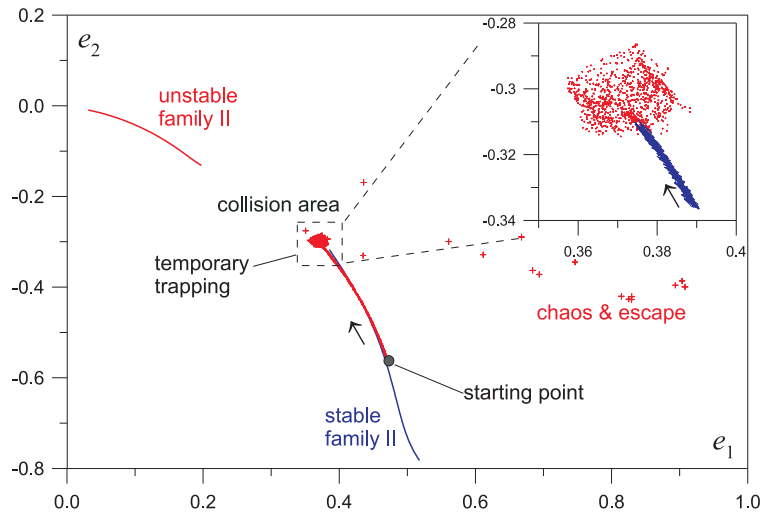}
\end{center}
\caption{The evolution of the system on the family II of the 2/1 resonant periodic orbits (see Fig. \ref{fam21-1}a). The starting point is at high eccentricities. The eccentricities decrease and the system is trapped in a bounded chaotic region close to the collision area, before  widespread chaos appears. A detail of the evolution close to the temporary chaotic trapping is shown in the separate plot.}
\label{fam21-2}
\end{figure}

In the previous study, we considered a system close to the family I of 2/1 resonant periodic orbits of Fig. \ref{fam21-1}a. We study now a system that starts close to the family II in this figure. The evolution is shown in Fig. \ref{fam21-2}. The starting point is close to $e_1=0.45$, $e_2=0.6$ on the family II. The system moves along the family, with decreasing eccentricities, until it is trapped temporarily at a point on the family, with smaller eccentricities, close to the collision area. The detail of this trapping is shown in the separate plot of Fig. \ref{fam21-2}. However, after this temporary capture, the system is disrupted due to the existence of strong chaos nearby the collision region.

\subsubsection{The inner planet has larger mass than the outer planet}

The fact that the inner planet is more massive than the outer planet results to an important qualitative difference for the family I of resonant periodic orbits: A small unstable region appears when $m_1\geq m_2$, which increases as the ratio $m_1/m_2$ increases (see Beaug\'e et al, 2006; Voyatzis et al, 2009). For both critical points formed, there bifurcate two asymmetric families of 2/1 resonant periodic orbits  In the following we use the masses $m_1=0.0015$, $m_2=0.0010$. For these values the two bifurcating asymmetric families meet and form a single family.  

\begin{figure}
\begin{center}
\includegraphics[width=6.5cm]{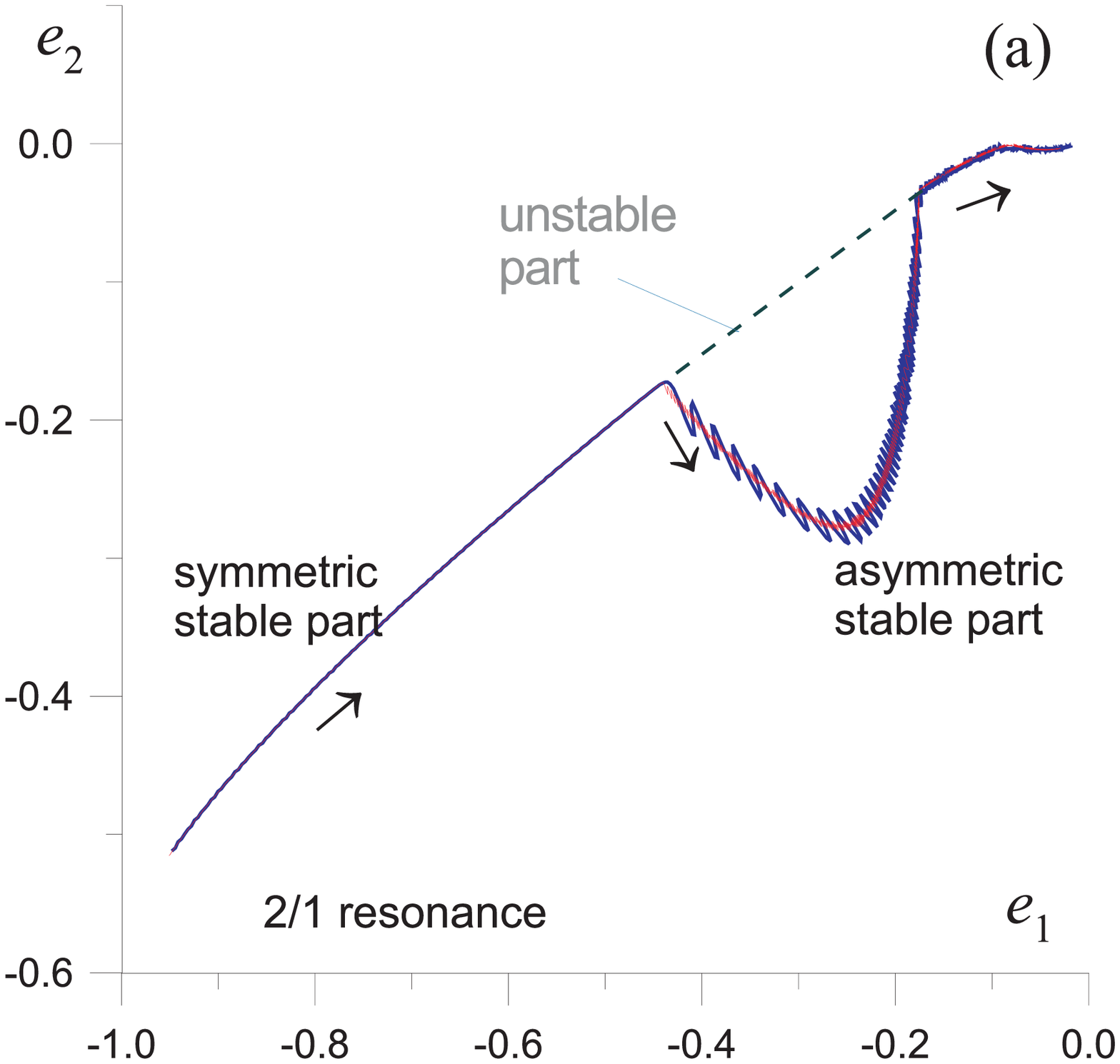}\hspace{0.5cm}
\includegraphics[width=6.5cm]{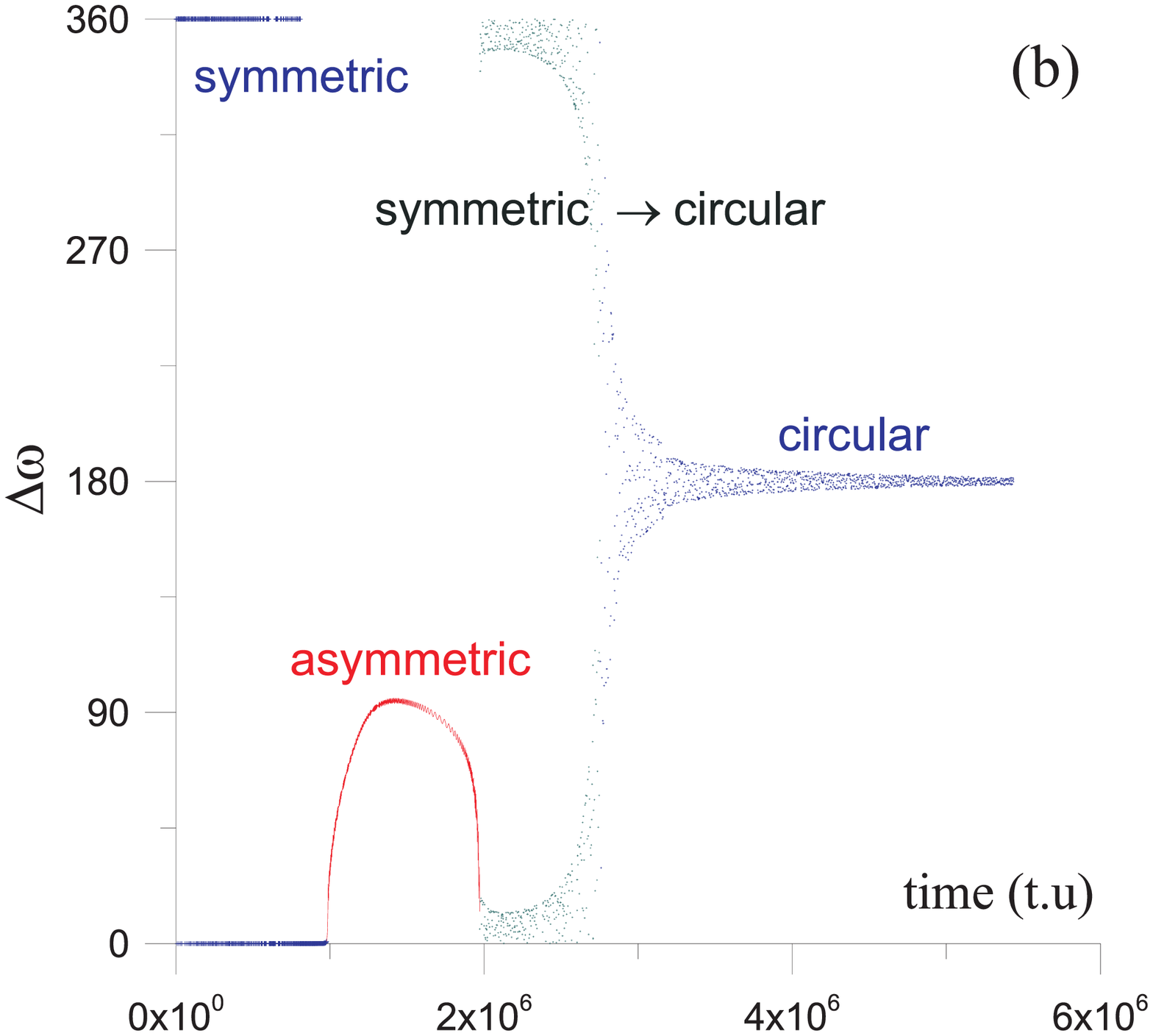}
\end{center}
\caption{(a) The evolution along the 2:1 resonant family of periodic orbits for $m_1=0.0015$, $m_2=0.0010$, starting from high eccentricities. The evolution follows the {\it stable} parts of the family. (b) The evolution of $\Delta\omega$, which indicates the passage of the system through different symmetric and asymmetric configurations.}
\label{fam21-3}
\end{figure}

In Fig. \ref{fam21-3}a we study the evolution starting from a point on the symmetric family, with high eccentricities. The system moves along the stable symmetric family until it meets the unstable region, where it changes route and follows the asymmetric family until this latter family meets again the symmetric family. From that point on it follows the path of the stable symmetric family and is finally trapped at a 2/1 resonant periodic orbit with zero eccentricities. The evolution is qualitatively similar for $\nu=5$ and $\nu=6$ in the dissipation law (\ref{diss1}).
In Fig. \ref{fam21-3}b we present the evolution of the angle $\Delta\omega$ between the line of apsides.  The transition from  symmetry $\Delta\omega=0$ to asymmetry ($\Delta\omega\neq 0^0, 180^\circ$) and again to symmetry, $\Delta\omega=0$ and then to $\Delta\omega=180^\circ$, is clearly seen.

\subsection{The dissipation law $\vec{R}=-10^{-\nu}(\vec{v}-\vec{v_c}/r^{1/2})$}

\begin{figure}
\begin{center}
\includegraphics[width=6.5cm]{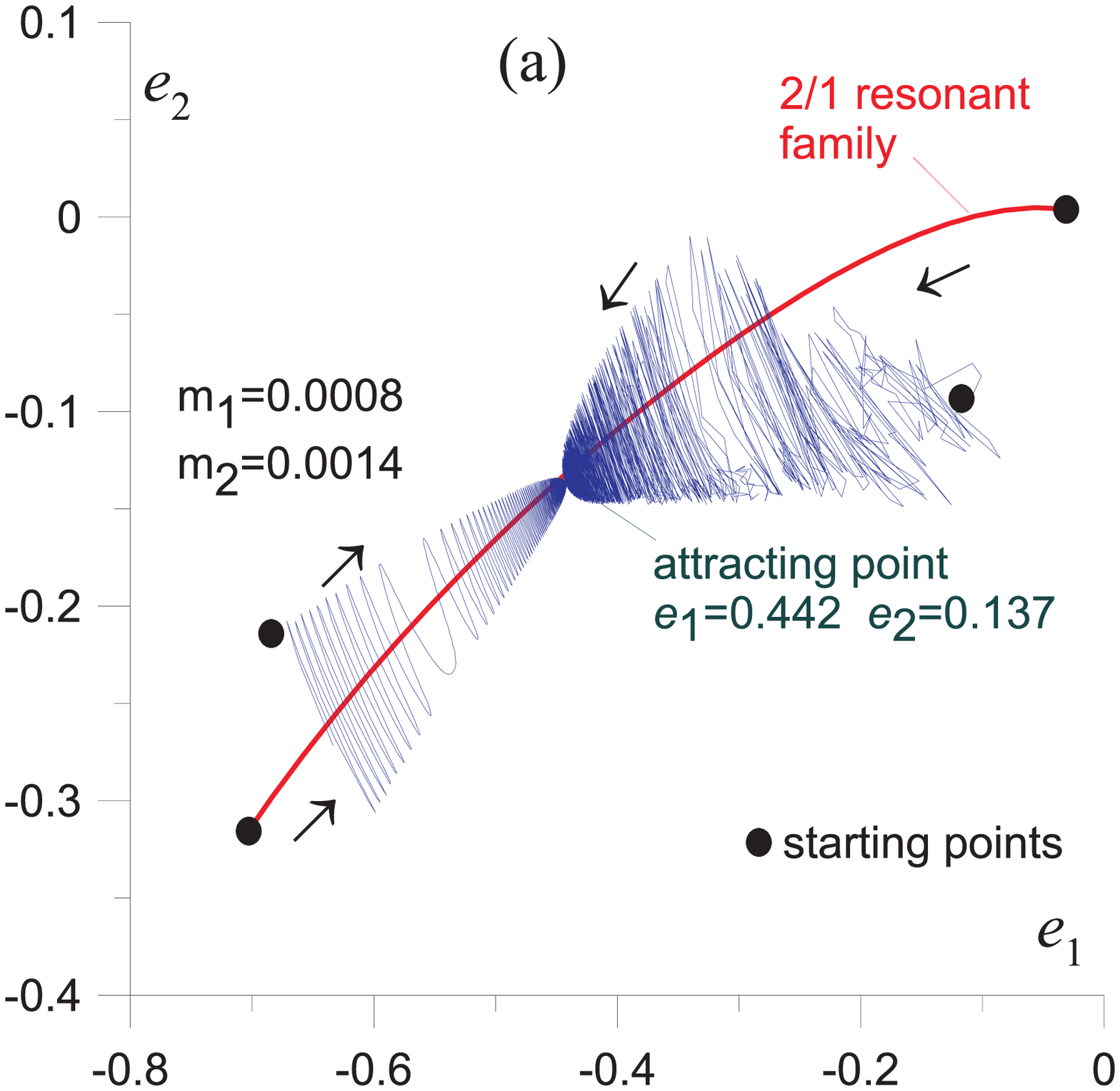}\hspace{1cm}
\includegraphics[width=6.5cm]{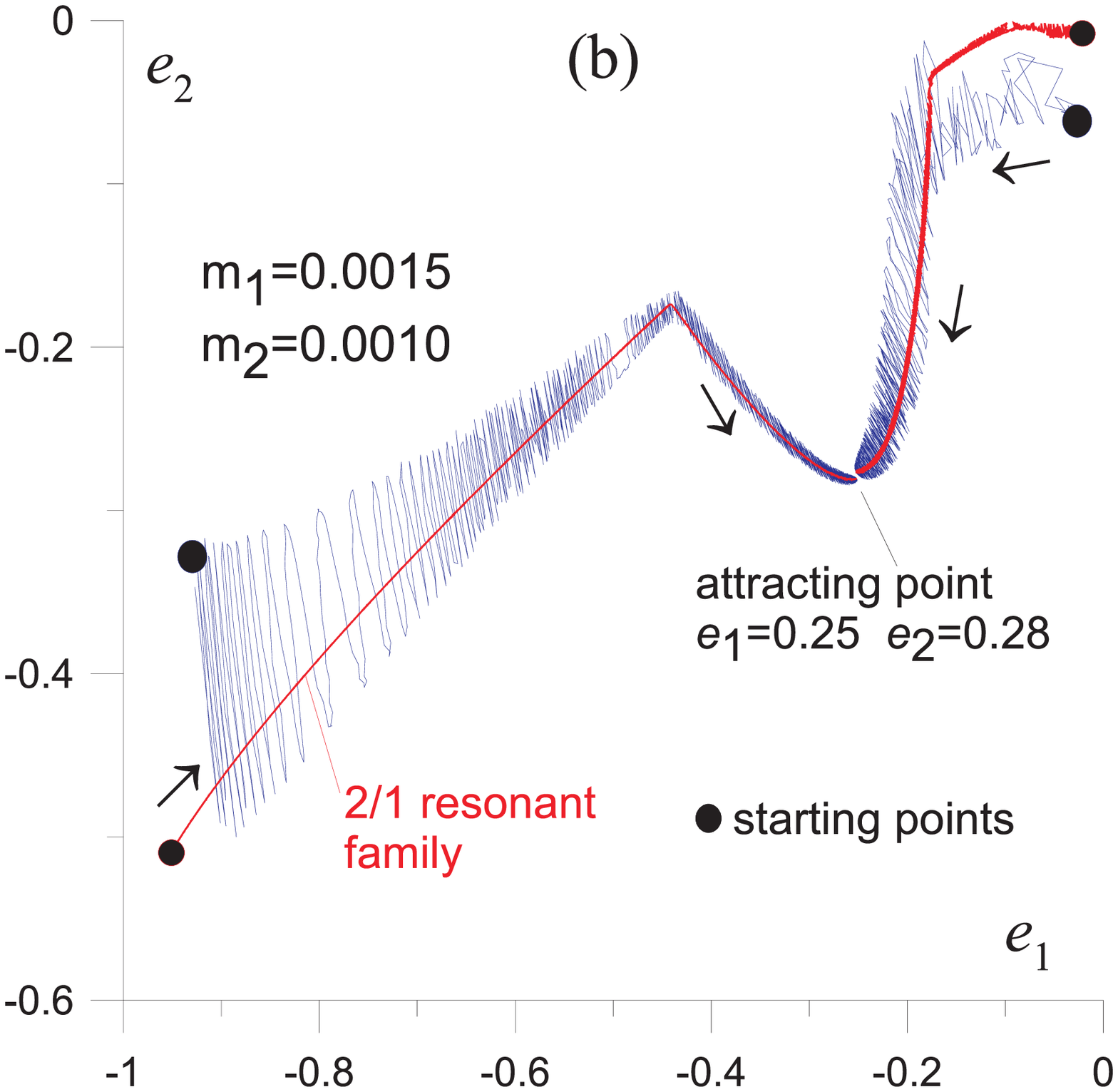}
\end{center}
\caption{The evolution of the system (in the eccentricity space) under the dissipation law $\vec{R}=-10^{-6}(\vec{v}-\vec{v_c}/r^{1/2})$, along a 2/1 resonant family, starting {\it on} the family and {\it close} to the family, both from small and from large eccentricities (absolute values). The system is trapped to a 2/1 resonant periodic orbit. (a) The evolution close to the family I for $m_1<m_2$ (see Fig. \ref{fam21-1}a). (b) Similar evolution as in panel (a), close to the family I for $m_1>m_2$ (see Fig. \ref{fam21-3}a).}
\label{e12-2-21}
\end{figure}

In this section we consider the dissipation law (\ref{diss2}) and repeat the same study as in section \ref{Sec21Res}, starting from two orbits on the family of 2/1 resonant periodic orbits, one with large eccentricities and the other with small eccentricities. We consider two cases: (1) $m_1<m_2$, using the masses $m_1=0.0008$, $m_2=0.0014$ and  (2) $m_1>m_2$, using the masses $m_1=0.0015$, $m_2=0.0010$. In case 1, the family of periodic orbits is given in Fig. \ref{fam21-1}a, family I, and in case 2 the family of periodic orbits is given in Fig. \ref{fam21-3}a.

The evolution is very different from that of Fig. \ref{fam21-1}. Starting from large eccentricities, the evolution is towards smaller eccentricities and starting from small eccentricities, the evolution is towards larger eccentricities. In both cases, the system is trapped to a 2/1 resonant periodic orbit on the corresponding family. The results are presented in Fig. \ref{e12-2-21}. In panel (a) we show the evolution when we start close to the family I of Fig. \ref{fam21-1}a. We start both with large and with small eccentricities (absolute values), and in each case we consider two starting points, one {\it on} the resonant family and one close to it. The system is attracted to the same orbit on this family, with $e_1=0,442$, $e_2=0.137$. If we start on the family, the system moves always on the family. If we start close to the family, then the system oscillates around the resonant family of periodic orbits, with decreasing amplitude, and is finally trapped to the periodic orbit mentioned above. We remark that the attracting orbit of Fig. \ref{e12-2-21}a is the same as the attracting orbit of Fig. \ref{tres-35}. In panel (b) we do the same work, starting close to the family of Fig. \ref{fam21-3}a. In this latter case, it is $m_1>m_2$ and the stable family includes an asymmetric branch. The evolution is the same as before, moving always along the {\it stable} family. In the present case the orbit to which the system is trapped is an orbit on the {\it asymmetric} branch of the 2/1 resonant periodic orbits.

\section{The evolution close to 3/2 resonant orbits under dissipation}

\begin{figure}
\begin{center}
\includegraphics[width=16cm]{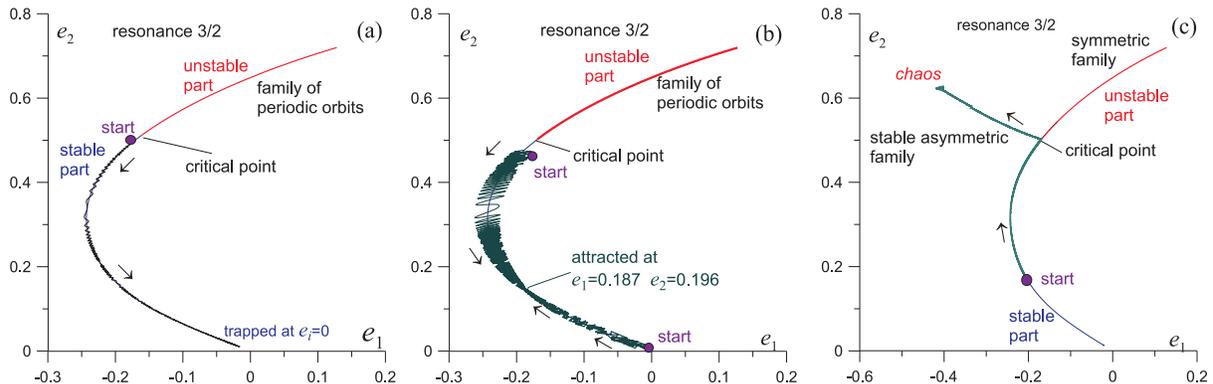}
\end{center}
\caption{The evolution close to the 3/2 resonance, under three different dissipation laws. (a) $\vec{R}=-10^{-6}(\vec{v}-\vec{v_c})$. (b) $\vec{R}=-10^{-6}(\vec{v}-\vec{v_c}/r^{1/2})$ (c) $\vec{R}=-10^{-7}(\vec{v}-\vec{v_c}/r^{3/2})$. The starting point and the direction of the evolution along the family of periodic orbits are indicated in all cases.}
\label{e12-32}
\end{figure}

In this section we present the evolution close to the 3/2 resonance, under three different dissipation laws. The results are presented in Fig. \ref{e12-32}a,b,c. In all panels the family of 3/2 resonant symmetric periodic orbits is presented, in the $e_1\:e_2$ space (the same family in the plane $x_1\:x_2$ is presented in Fig. \ref{families}a). Along the family, the stability changes and after a certain point instability appears. At the critical point, a family of 3/2 resonant stable asymmetric periodic orbits bifurcates (Michtchenko et al, 2006). In all cases we used the planetary masses $m_1=0.0008$, $m_2=0.0014$.

In panel (a) we study the evolution under the law (\ref{diss1}), particularly $\vec{R}=-10^{-6}(\vec{v}-\vec{v_c})$. The starting point is at large eccentricities and the system evolves to zero eccentricities. This evolution is similar to that obtained for the 2/1 and 3/1 resonance but now the semimajor axes decrease slightly at first and then they are stabilized to fixed values and the system is trapped in a particular circular orbit.

In panel (b) we present the evolution of the system under the law (\ref{diss2}), particularly $\vec{R}=-10^{-6}(\vec{v}-\vec{v_c}/r^{1/2})$. We start from two points, one with large and one with small eccentricities, as in the case of Fig. \ref{e12-2-21}a. The evolution is similar as in this latter figure: We find that the system is attracted to a periodic orbit with $e_1=0.187$, $e_2=0.196$, on the 3/2 resonant family, by decreasing or increasing eccentricities, respectively. 

In panel (c) we study the evolution under the law $\vec{R}=-10^{-7}(\vec{v}-\vec{v_c}/r^{3/2})$, which differs from the law used in panel (b). In this case the velocity of rotation of the protoplanetary nebula is stronger than the Keplerian circular velocity $\vec v_c$ close to the sun and decreases more rapidly for larger distances, compared to the dissipation law of panel (b). We start from small eccentricities and we note that the eccentricities increase, until the critical point where instability starts. From that point on, it follows the stable 3/2 asymmetric family, up to a certain point before chaos appears and the system is disrupted.

\section{Discussion}

In this work we studied the evolution of a planetary system with two planets moving in the same plane, under the action of dissipative forces. The main study was made for the Stokes like force given by Eq. (\ref{diss1}), for different values of $\nu$. This force implies that the two planets move inside a protoplanetary nebula that rotates with a differential Keplerian circular velocity $\vec v_c$. We also considered a variation of this dissipative force given by eq. (\ref{diss2}) and, for the 3/2 resonance, also the force $\vec{R}=-10^{-\nu}(\vec{v}-\vec{v_c}/r^{3/2})$. In these last cases the rotational velocity of the protoplanetary nebula deviates from the circular Keplerian velocity, being larger close to the sun and tending to zero at large distances. 

An important result that was obtained in the present study is that the families of periodic orbits play a crucial role in the evolution of the system: These families, both stable or unstable, determine the structure of the phase space. It was shown that the {\it stable} families provide the routes along which the system evolves, while the unstable families define the regions that the system avoids in its evolution. {\em The evolution may be different for different dissipation laws, but in all cases it is the families of periodic orbits that guide the system in its evolution, either to smaller or to larger eccentricities}. The particular dissipation law determines how the system moves along the families and its final trapping to a resonance.

We studied the evolution of a planetary system considering two different cases. 

Case 1: The planetary system starts far from a periodic motion (either resonant or non resonant).\\
Case 2: The planetary system starts close to a resonant periodic motion. In particular, we studied the 2/1 and 3/2 resonances.

The evolution depends on the particular dissipative force. 

\underline{Case 1}:\\ 
We considered first the Stokes like law of eq. (\ref{diss1}). We found that in all cases the system is attracted to an orbit on the {\it circular family of periodic orbits}, following an irregular route with almost the same value of the planetary frequencies $n_1/n_2$ as the starting value. After this, the system evolves {\it along} the circular family, in a smooth way, to the direction where $n_1/n_2$ decreases, until it meets a significant resonance, for example 2/1 or 3/1. From that point on, the system is trapped in a chaotic attractor or in a resonant orbit, respectively, with very small eccentricities (see figures \ref{e12-28}, \ref{e12-35}). The semimajor axes increase slowly and then are stabilized at the values corresponding to the resonant attractor. This evolution is the same for a whole region in the eccentricity space, which is in fact the {\it basin of attraction} to the above mentioned chaotic attractor (Fig. \ref{e12-35-all}) for the 3/1 resonance or to a circular resonant orbit for the 2/1 resonance. Thus we get a mechanism of trapping to a resonance with very small eccentricities.

A different evolution appears for the dissipation law (\ref{diss2}). The system moves towards a 3/1 resonant orbit on the circular family and then follows the {\it stable} family of elliptic 3/1 resonant periodic orbits that bifurcates from the resonant circular orbit. It evolves along this family, with {\it increasing} eccentricities, and if the family becomes unstable, it changes route and follows the stable asymmetric family that bifurcates from this critical point. This process provides a mechanism of trapping to an asymmetric 3/1 resonant periodic orbit with large eccentricities.

\underline{Case 2}:\\
We start with the Stokes like eq. (\ref{diss1}), as in case 1. We study first the evolution close to the 2/1 resonance. In all cases, the evolution follows the {\it stable} parts of the family of 2/1 resonant periodic orbits, with {\it decreasing} eccentricities (see figures \ref{fam21-1}b, \ref{fam21-3}b for the 2/1 resonance and Fig. \ref{e12-32}a for the 3/2 resonance). The semimajor axes are stabilized to fixed values, after a slow decrease. If, however, there is a chaotic region along the family, then the system develops chaotic motion (unless the dissipation process stops earlier) with decreasing eccentricities (Fig. \ref{fam21-2}). This evolution is the same either for a starting orbit on the 2/1 family or far from it.

The evolution is different for the dissipative force given by eq. (\ref{diss2}). Depending on the starting point, the eccentricities decrease or increase and the system is finally trapped to a resonant periodic orbit with nonzero eccentricities (see Fig.\ref{e12-2-21} for the 2/1 resonance and Fig. \ref{e12-32}b for the 3/2 resonance).

Finally, a different evolution appears for the dissipative force $\vec{R}=-10^{-\nu}(\vec{v}-\vec{v_c}/r^{3/2})$. Contrary to the previous two cases, the eccentricities in this case {\it increase}, following the stable parts of the resonant family, Chaotic motion may appear if an unstable region is met on the family (see Fig.\ref{e12-32}c for the 3/2 resonance).

We see that the final outcome, whether the system is trapped to small or large eccentricities, depends critically on the particular dissipation force, and in particular on how much the rotational velocity of the protoplanetary nebula deviates from the Keplerian circular velocity. In almost all cases the system is trapped to a resonant orbit.


\begin{thebibliography}{}

\bibitem{Baluev}
Baluev R. V. : Resonances of low order in the planetary system HD37124, Cel.Mech.Dyn.Astron., {\bf 102}, 297--325 (2008).

\bibitem{Beauge93}
Beaug\'e, C., Ferraz-Mello, S.: Resonance trapping in the primordial solar nebula: the case of Stokes drag dissipation, Icarus {\bf 103}, 301--318 (1993).

\bibitem{Beauge03}
Beaug\'e, C., Ferraz-Mello, S. and Michtchenko, T. : Extrasolar Planets in Mean-Motion Resonance: Apses Alignment and Asymmetric Stationary Solutions,  ApJ {\bf 593}, 1124--1133 (2003).

\bibitem{Beauge06}
Beaug\'e C., Ferraz-Mello  S., Michtchenko T.A. : Planetary Migration and Extrasolar Planets in the 2/1 Mean-motion Resonance, MNRAS {\bf 365}, 1160--1170 (2006).

\bibitem{Beauge08}
Beaug\'{e} C., Giuppone C.A., Ferraz-Mello S., Michtchenko T.: Reliability of orbital fits for resonant extrasolar planetary systems: the case of HD82943, MNRAS {\bf 385}, 2151--2160 (2008).

\bibitem{Calleg06}
Callegari N., Feraraz-Mello S., Michtchenko T.A. : Dynamics of two planets in the 3/2 resonance: application to the planetary system of the pulsar PSR B1257+12, Cel.Mech.Dyn.Astron. {\bf 94}, 381--397 (2006).

\bibitem{Contopoulos}
Contopoulos G. : {\em Order and Chaos in Dynamical Astronomy}, Springer, Berlin 2002.

\bibitem{SFM08}
Ferraz-Mello, S., Rodriguez A. : Tidal friction in close-in satellites and exoplanets:the Darwin theory re-vised, Cel.Mech.Dyn.Astron. {\bf 101}, 171--201 (2008).

\bibitem{SFM03}
Ferraz-Mello, S., Beaug\'e, C. and Michtchenko T.A. : Evolution of migrating planet pairs in resonance,  Cel.Mech.Dyn.Astron. {\bf 87}, 99--112 (2003).

\bibitem{06} 
Ferraz-Mello S., Michtchenko, T.A., Beaug\'e C. : Regular motion in extrasolar planetary systems, in B.A. Steves, A.J. Maciejewski and M.Henry (eds.), `Chaotic Worlds: From Order to Disorder in Gravitational N-body Dynamical Systems', Springer--NATO Science Series, {\bf 227}, 255--288 (2006)

\bibitem{Gomes96}
Gomes R.S. : The effect of Nonconservative forces on resonance lock: stability and instability, Icarus {\bf 115}, 47--59 (1996).

\bibitem{Godj08}
Gozdziewski K., Migaszewski C., Konacki M.: A dynamical analysis of the 14 Herculis planetary system, MNRAS {\bf 385}, 957--966 (2008).

\bibitem{Hadjidem75}
Hadjidemetriou J.D.: The continuation of periodic orbits from the restricted to the general three-body problem, Cel. Mech. {\bf 12}, 155--174 (1975).

\bibitem{Hadjidem02}
Hadjidemetriou, J.D. : Resonant periodic motion and the stability of extrasolar planetary systems,  Cel.Mech.Dyn.Astron. {\bf 83}, 141--154 (2002).

\bibitem{Hadjidem06}
Hadjidemetriou, J.D. : Symmetric and Asymmetric Librations in Extrasolar Planetary Systems: A global view, Cel.Mech.Dyn.Astron. {\bf 95}, 225--244 (2006).

\bibitem{Hadjidem08}
Hadjidemetriou, J.D. : On periodic orbits and resonance in extrasolar planetary systems, Cel.Mech.Dyn.Astron. {\bf 102}, 69--82 (2008)

\bibitem{Hanghi99}
Haghighipour N. : Dynamical friction and resonance trapping in planetary systems, MNRAS {\bf 304}, 185--194 (1999).

\bibitem{Lee02}
Lee, M.H. and Peale, S. : Dynamic and origin of the 2:1 orbital resonances of the GJ 876 planets,  ApJ {\bf 567}, 596--609 (2002).

\bibitem{Lee04}
Lee, M. H. : Diversity and Origin of 2:1 Orbital Resonance in Extrasolar Planetary Systems,  ApJ {\bf 611}, 517--527 (2004). 

\bibitem{Marzari06}
Marzari F., Scholl H., Tricarico P. : A numerical study of the 2:1 planetary resonance, A\&A {\bf 453}, 341--348 (2006).

\bibitem{Micht06}
Michtchenko T.A., Beaug\'e C., Ferraz-Mello S.: Stationary Orbits in Resonant Extrasolar Planetary Systems, Cel.Mech.Dyn.Astron. {\bf 94}, 381--397 (2006). 

\bibitem{Morbi07}
Morbidelli A., Tsiganis K., Grida A., Levison H.F., Gomes R. : Dynamics of the giant planets of the solar system in the gaseous protoplanetary disk and their relation to the current orbital architecture, Astron. J. {\bf 134}, 1790--1798 (2007).

\bibitem{Nelson03a}
Nelson R.P., Papaloizou J.C.B. : The interaction of a giant planet with a disk with MHD turbulence - I. The initial turbulent disc models, MNRAS {\bf 339}, 983--992 (2003a)

\bibitem{Nelson03b}
Nelson R.P., Papaloizou J.C.B. : The interaction of a giant planet with a disk with MHD turbulence - II. The interaction of the planet with the disk, MNRAS {\bf 339}, 993--1005 (2003b)

\bibitem{Papalo03}
Papaloizou J.C.B. : Disk planet interactions: migration and resonances in extrasolar systems, Cel.Mech.Dyn.Astron. {\bf 87}, 53--83 (2003).

\bibitem{Phyhoyos05}
Psychoyos, D. and Hadjidemetriou, J.D. : Dynamics of 2/1 resonant extrasolar systems. Application to HD82943 and Gliese876, Cel.Mech.Dyn.Astron. {\bf 92}, 135--156 (2005).

\bibitem{Sandor06}
Sandor Zs., Kley.W. : On the evolution of the resonant planetary system HD128311, A\&A {\bf 451}, 31--34 (2006).

\bibitem{Schneider}
Schneider J. : `The Extrasolar Planets Encyclopaedia', http://exoplanet.eu/ (2009).

\bibitem{Schwrz09}
Schwarz R., Suli A., Dvorak R. and Pilat-Lohinger, E. : Stability of Trojan planets in multi-planetary systems. Stability of Trojan planets in different dynamical systems, Cel.Mech.Dyn.Astron. {\bf 104}, 69--84 (2009).  

\bibitem{Tsiga05}
Tsiganis K., Gomes R., Morbidelli A., Levison H.F. : Origin of the orbital architecture of the giant planets of the Solar system, Nature {\bf 435}, 459--461 (2005).

\bibitem{Voya05}
Voyatzis G., Hadjidemetriou H.D. : Symmetric and asymmetric librations in planetary and satellite systems at the 2/1 resonance, Cel.Mech.Dyn.Astron., {\bf 93}, 263--294 (2005).

\bibitem{Voya06}
Voyatzis, G. and Hadjidemetriou , J.D. : Symmetric and asymmetric 3/1 resonant periodic orbits: An application to the 55Cnc extra-solar system,  Cel.Mech.Dyn.Astron. {\bf 95}, 259--271 (2006). 

\bibitem{Voya08}
Voyatzis G. : Chaos, order and periodic orbits in 3:1 resonant planetary dynamics, ApJ {\bf 675}, 802--816 (2008).

\bibitem{Voya09}
Voyatzis G., Kotoulas T., Hadjidemetriou J.D. : On the 2/1 resonant planetary systems - periodic orbits and dynamical stability, MNRAS {\bf 395}, 2147--2156 (2009).

\bibitem{Zhou08}
Zhou L.Y., Ferraz-Mello S., Sun Y.S. : Formation of the 3:1 MMR in 55CnC system, in I.A.U. Proc. No.249, `Exoplanets: Detection Formation and Dynamics', Y.S.Sun, S.Ferraz-Mello and J.L.Zhou (eds), 119--124 (2008). 

\end{thebibliography}
\end{document}